\documentclass[reprint,two-column,double-spaced,groupedaddress,superscriptaddress,pra]{revtex4-2}

\usepackage[normalem]{ulem}

\usepackage{mathrsfs}
\usepackage{enumerate}
\usepackage{graphicx}
\usepackage{physics}

\usepackage[colorlinks]{hyperref}
\usepackage{tensor}
\usepackage{leftidx}
\usepackage{braket}

\begin{document}


\title{Identifying genuine quantum teleportation}


\author{Chia-Kuo Chen}
\thanks{These authors contributed equally to this work.}
\affiliation{Department of Engineering Science, National Cheng Kung University, Tainan 70101, Taiwan}
\affiliation{Center for Quantum Frontiers of Research $\&$ Technology, National Cheng Kung University, Tainan 70101, Taiwan}

\author{Shih-Hsuan Chen}
\thanks{These authors contributed equally to this work.}
\affiliation{Department of Engineering Science, National Cheng Kung University, Tainan 70101, Taiwan}
\affiliation{Center for Quantum Frontiers of Research $\&$ Technology, National Cheng Kung University, Tainan 70101, Taiwan}

\author{Ni-Ni Huang}
\affiliation{Department of Engineering Science, National Cheng Kung University, Tainan 70101, Taiwan}
\affiliation{Center for Quantum Frontiers of Research $\&$ Technology, National Cheng Kung University, Tainan 70101, Taiwan}

\author{Che-Ming Li}
\email{cmli@mail.ncku.edu.tw}
\affiliation{Department of Engineering Science, National Cheng Kung University, Tainan 70101, Taiwan}
\affiliation{Center for Quantum Frontiers of Research $\&$ Technology, National Cheng Kung University, Tainan 70101, Taiwan}
\affiliation{Center for Quantum Technology, Hsinchu 30013, Taiwan}



\date{\today}

\begin{abstract}
An unknown quantum state can be teleported by using quantum measurements and the maximally entangled Einstein-Podolsky-Rosen (EPR) pair. It is well known that the usual nonclassical teleportation that cannot be simulated by the seminal classical measure-prepare strategy can be demonstrated with all entangled states. Herein, we propose a new benchmark which reveals that not all such nonclassical teleportations are truly quantum-mechanical. Rather, there exists a more robust classical-teleportation model, which includes the measure-prepare mimicry as a special case, that can describe certain nonclassical teleportations. Invalidating such a general classical model indicates \emph{genuine quantum teleportation} wherein both the pair state and the measurement are truly quantum-mechanical. We prove that EPR steering empowers genuine quantum teleportations, rather than entanglement. The new benchmark can be readily used in practical experiments for ensuring that genuine quantum teleportation is implemented. The results presented herein provide strict criteria for implementing quantum-information processing where genuine quantum teleportation is indispensable.
\end{abstract}


\maketitle

\section{INTRODUCTION}

\begin{figure*}[t]
\includegraphics[width=17.8cm]{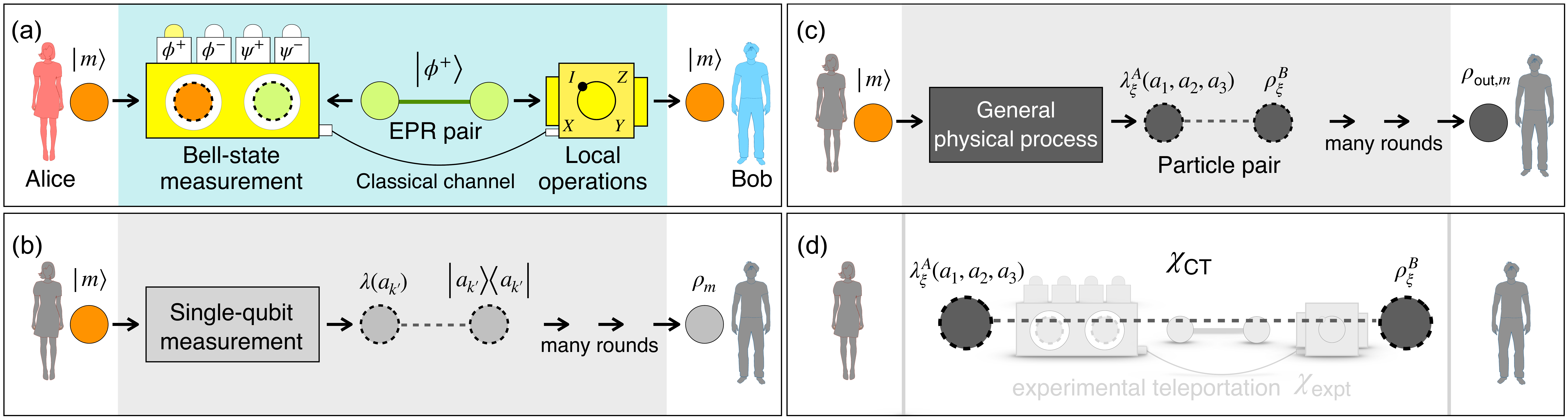}
\caption{Quantum and classical teleportation. (a) Quantum teleportation. The sender, Alice, and, receiver, Bob first share an Einstein-Podolsky-Rosen (EPR) pair. Alice performs Bell-state measurement on the unknown transmitted qubit (with input state $\ket{m}$) and half of the entangled pair held by her in the basis of Bell states: $\{\ket{\phi^+},\ket{\phi^-},\ket{\psi^+},\ket{\psi^-}\}$, where $\ket{\phi^\pm}\equiv(\ket{00}\pm\ket{11})/{\sqrt{2}}$ and $\ket{\psi^\pm}\equiv(\ket{01}\pm\ket{10})/{\sqrt{2}}$. She then sends her measurement result to Bob over a classical channel. Depending on Alice's outcome, Bob performs the local Pauli operations $I,X,Y,$ and $Z$ on his half of the entangled pair to recover the unknown state. (b) Measure-prepare strategy. Alice directly measures $\ket{m}$ for an observable $A_{k'}$ and then sends the result $a_{k'}$ to Bob to prepare the corresponding state $\ket{a_{k'}}\!\!\bra{a_{k'}}$ as the output state. Alice's measurements make the input states classical in the sense that the system is composed of pre-existing states, $\lambda(a_{k'})$. After many rounds of measurements, Alice obtains the average output state $\rho_{m}$ (see Eq.~(\ref{rm}) with $p_{k'}=1$). (c) Generic classical teleportation. The input state $\ket{m}$ undergoes a generic physical process and decays into a classical state that can be described by a pre-existing state $\lambda^{A}_{\xi}(a_1,a_2,a_3)$. The state $\lambda^{A}_{\xi}$ correlates to the output state of Bob's particle, $\rho^{B}_{\xi}$. The final states, $\rho_{\rm{out},m}$, created from the mixture of the states, $[\lambda^{A}_{\xi};\rho^{B}_{\xi}]$, are used to construct the process matrix, $\chi_{\rm{CT}}$. See Eqs.~(\ref{rhoQPT}) and~(\ref{ct}) for details. (d) Utility of the generic classical teleportation ($\chi_{\rm{CT}}$). Since such a process reveals the extent to which teleportation can be performed with the classical particle pairs of states $[\lambda^{A}_{\xi};\rho^{B}_{\xi}]$ ({c}), it provides a classical-teleportation model by which experiments ($\chi_{\rm{expt}}$) can be quantified to show the amounts of genuine quantum teleportation and classical teleportation.}
\label{QT}
\end{figure*}

Transmitting the unknown state of a quantum system from a sender to a remote receiver without directly sending the system itself is made possible by \textit{quantum teleportation} \cite{Bennett93}. The procedure of teleporting quantum states exploits both quantum measurements \cite{Braunstein95} and the quantum correlation of the Einstein-Podolsky-Rosen (EPR) pair \cite{Horodecki09}. See Fig.~\ref{QT}(a). Quantum teleportation is a novel means of communication that has no classical analogues, and moreover, provides a new method for coherently manipulating quantum states. Efforts to explore the possibilities provided by teleportation have led to the development of many feasible techniques for engineering quantum systems \cite{Bouwmeester97,Nielsen98,Furusawa98,Barrett04,Sherson06,Xia18} and have facilitated a deeper understanding of quantum information \cite{Bennett00,Dowling03,Pirandola15}.

Quantum teleportation constitutes the essential elements required to perform a range of quantum computation and quantum-information tasks. In particular, to construct a large-scale quantum computing processor with a quantum modular architecture \cite{Monroe13,Devoret13}, teleportation is needed to integrate the various modules \cite{Eisert00,Jiang07}. Furthermore, teleportation can be extended to send controlled gates across different modules of a quantum processor to realize universal quantum computation \cite{Gottesman99,Chou18}. Finally, teleportation is essential for realizing modular architectures of quantum networks by which spatially separated quantum nodes can communicate with each other \cite{Pirker18}. While there exist many enabling advances for quantum teleportation, ideal teleportation is considered to be the default therein. Identifying quantum teleportation for real-world implementation~\cite{Pirandola15,Popescu94,Badziag00,Duan00,Verstraete03,Buono12,He15,Hsieh17,Carvacho18,Chen20,Huang20}, therefore, is not only significant in its own right, but also fundamental in releasing the true power of quantum-information processing.

The measure-prepare strategy~\cite{Pirandola15,Massar95} offers a seminal paradigm for transmitting unknown quantum states in the absence of EPR pairs. In particular, the unknown quantum states are measured by the sender to infer what the input must have been, and the receiver then prepares the output state accordingly. Evaluating whether an experimental teleportation can outperform this mimicry to serve as a so-called {\it nonclassical teleportation} has become a widely-used reliability standard for teleportation experiments \cite{Ma12,Ren17,Bao12,Olmschenk09,Steffen13,Pfaff14,Takeda13}. Such an evaluation has also led to significant research on what physical resources are actually required to achieve nonclassical teleportation \cite{Horodecki99}. For example, it was recently reported by Cavalcanti {\it et al}. \cite{Cavalcanti17} that, once the pair shared between the sender and receiver is entangled, nonclassical teleportation can be demonstrated under partial Bell-state measurement (BSM).

Driven by the desire to ultimately identify quantum teleportation, we consider herein the following fundamental question:\\

\noindent{\it To what extent can teleportation be performed without using the quantum resource of EPR pairs?}\\

\noindent The answer has a profound effect on how quantum teleportation can be faithfully realized, and leads naturally to the next question:\\

\noindent{\it Can all entangled states enable genuine quantum teleportation that outperforms such general classical teleportation?}\\

\noindent To address these questions, we begin by introducing a novel classical-teleportation model to simulate quantum teleportation without the use of EPR pairs. We further introduce a new benchmark consisting of fidelity criteria and quantitative identifications for unambiguously determining the extent to which classical methods can optimally mimic quantum teleportation. We show that measure-prepare mimicry \cite{Massar95}, including the extended version introduced by Cavalcanti {\it et al.} \cite{Cavalcanti17}, is a special case of this general classical-teleportation model. We then use these results to answer the second question, and show that EPR steering powers genuine quantum teleportation, rather than entanglement. Finally, we illustrate the application of our proposed formalism through several concrete examples, including its application to present teleportation experiments. We demonstrate that the standard criterion for surpassing the maximum average fidelity between the teleported states and target states that can be achieved using measure-prepare mimicry, i.e., $\bar{F}_{\rm{expt},\rm{s}}>2/3\sim0.667$ \cite{Massar95,Cavalcanti17}, does not guarantee the existence of genuine quantum teleportation that outperforms the introduced classical teleportation. We further show that a stricter fidelity criterion of $\bar{F}_{\rm{expt},\rm{s}}>0.789$ is required for implementing such high-quality teleportation.

\section{Characterizing teleportation processes}
In order to make the proposed framework amenable to a wide variety of circumstances and applications in practical experiments \cite{Xia18,Pirandola15,Ma12,Bao12,Olmschenk09,Steffen13}, we systematically exploit experimentally measurable quantities and then completely characterize the teleportation process using quantum process tomography (QPT) \cite{Chuang97,Nielsen00}. Moreover, we consider a general scenario wherein there exists a verifier, Victor, who follows a specific protocol to examine experimental teleportation \cite{Pirandola15}. We assume that Victor randomly provides Alice with the following four states as input states of teleportation: $\ket{0}$, $\ket{1}$, $\ket{+}$, and $\ket{R}$, where $\ket{+}=(\ket{0}+\ket{1})/\sqrt{2}$, $\ket{R}=(\ket{0}+i\ket{1})/\sqrt{2}$, and $\ket{0}$ and $\ket{1}$ constitute an orthonormal basis.
Since the states are randomly chosen by Victor, the input states are unknown to Alice and the receiver, Bob, in each round of the experimental teleportation.
Then, Victor receives the output states from Bob and performs state tomography \cite{Nielsen00} on the output states following the QPT protocol. After many runs of experiments, Victor can fully describe the teleportation process by a positive Hermitian process matrix, denoted as $\chi_{\rm{expt}}$. That is, $\chi_{\rm{expt}}$ is constructed from the output states, $\rho_{\rm{out},m}$, conditioned on the input states $\rho_{\rm{in},m}=\ket{m}\!\!\bra{m}$ for $m=0,1,+,$ and $R$, and has the form:
\begin{equation}
\chi_{\rm{expt}}
=\frac{1}{2}\left[\begin{array}{cc}\rho_{\rm{out},0} & \rho_{\rm{out},+}+i\rho_{\rm{out},R}-\tilde{I}_{\rm{out}} \\ \rho_{\rm{out},+}-i\rho_{\rm{out},R}-\tilde{I}_{\rm{out}}^{\dag} & \rho_{\rm{out},1}\end{array}\right],\label{exp}
\end{equation}
where $\tilde{I}_{\rm{out}}=e^{i\pi/4}(\rho_{\rm{out},0}+\rho_{\rm{out},1})/\sqrt{2}$. (See Appendix A for full derivation of Eq.~(\ref{exp}).) In the ideal case where $\rho_{\rm{out},m}=\rho_{\rm{in},m}$, ideal teleportation [Fig.~\ref{QT}(a)] is described by
\begin{equation}
\chi_{\rm{I}}\!=\frac{1}{2}\!\left[\begin{array}{cc}\ket{0}\!\!\bra{0} & \ket{+}\!\!\bra{+}\!+\!i\ket{R}\!\!\bra{R}\!-\!\hat{I} \\ \ket{+}\!\!\bra{+}\!-\!i\ket{R}\!\!\bra{R}\!-\!\hat{I}^{\dag} & \ket{1}\!\!\bra{1}\end{array}\right],\label{i}
\end{equation}
where $\hat{I}=e^{i\pi/4}I/\sqrt{2}$ and $I$ is the identity matrix.

It is worth noting that the manner in which the input system evolves from $\rho_{\rm{in},m}$ to $\rho_{\rm{out},m}$ can be specified by the process matrix $\chi_{\rm{expt}}$ through the mapping $\chi_{\rm{expt}}(\rho_{\rm{in},m})=\rho_{\rm{out},m}$, where this mapping preserves the Hermiticity, trace and positivity of the original density matrix of the system. Furthermore, $\chi_{\rm{expt}}$ is experimentally measurable if quantum state tomography \cite{Nielsen00} can be implemented with specific quantum measurements to obtain $\rho_{\rm{out},m}$. (See the practical measurement of $\chi_{\rm{expt}}$ in the existing experiments reported in \cite{Pirandola15,Ma12,Bao12,Olmschenk09,Steffen13}.) We will hereafter use a process matrix to refer to a teleportation process within the text.\\

\section{Implication of measure-prepare strategy}

To consider the extent to which teleportation can be implemented in the absence of EPR pairs, let us first revisit the basic ideas of the measure-prepare strategy \cite{Massar95} in mimicking quantum teleportation. In such a situation, Alice measures the unknown input state directly, and then sends the results via a classical communication channel to Bob to prepare the output state.
 Suppose that, in each measurement round, Alice chooses a physical property, $A_k$, for the measurement on the input state $\ket{m}$. Alice then obtains an outcome, ${a_k}$, where $a_k\in\{+1,-1\}$, from which Bob prepares the output state as the eigenstate, $\ket{a_k}$, of the observable for $A_{k}$. After many rounds of measurements under the same input state $\ket{m}$, and taking all of Bob's final states into account, the average output state from the measure-prepare procedure can be obtained as
 \begin{equation}
 \rho_{m}=\sum_{k}p_{k}\sum_{{a_k}}p_{{a_k},m}\ket{a_k}\!\!\bra{a_k},\label{rm}
 \end{equation}
 where $p_{k}$ is the probability of choosing the $k$th observable $A_{k}$ for measurement, and $p_{{a_k},m}=|\!\left\langle {a_k}|m\right\rangle\!|^{2}$ describes the probability of observing $a_{k}$ when the state being measured is $\ket{m}$ under the chosen $A_{k}$. The derivation of $\rho_{m}$ is shown in Appendix B.

 When averaged uniformly over all the possible input states $\ket{m}$, the average state fidelity of $\ket{m}$ and the output states $\rho_{m}$ is $\bar{F}_{\rm{expt},\rm{s}}=2/3$ \cite{Massar95}. Moreover, according to the relation between the average state fidelity, $\bar{F}_{\rm{expt},\rm{s}}$, and the process fidelity, $F_{\rm{expt}}\equiv tr(\chi_{\rm{I}}\chi_{\rm{expt}})$, by \cite{Gilchrist05}
 \begin{equation}
\bar{F}_{\rm{expt},\rm{s}}=\frac{1}{3}(2F_{\rm{expt}}+1),\label{fsp}
\end{equation}
where the process fidelity for the measure-prepare process is
 \begin{equation}
 F_{\rm{expt}}=\frac{1}{2}.\label{1/2}
 \end{equation}
 In other words, from the process viewpoint, the similarity between measure-prepare mimicry and ideal quantum teleportation is $50\%$.

Alice's measurement processes make the input states classical such that the output states $\rho_{m}$ can be described by classical realistic theory. For illustration purposes, let us assume that only one observable, say $A_{k'}$, is of interest. The average output state from the measure-prepare procedure, Eq.~(\ref{rm}), then becomes $\rho_{m}=\sum_{{a_{k'}}}p_{{a_{k'}},m}\ket{a_{k'}}\!\!\bra{a_{k'}}$. One can think of $\rho_{m}$ as an experimental output composed of pre-existing states, denoted by $\lambda(a_{k'})$, with a probability distribution $p_{{a_{k'}},m}$. In other words, once we have the pre-existing recipe of $\lambda(a_{k'})$ with the probability distribution $p_{{a_{k'}},m}$ for state preparation, the output state, $\rho_{m}$, can be prepared accordingly by incoherently mixing the states $\ket{a_{k'}}$. Obviously, $\lambda(a_{k'})$ exists independently of observation; after all, the prescribed information about $a_{k'}$ is already shown in the pre-existing recipe (see~Fig.~\ref{QT}(b)). This implication of the measure-prepare strategy motivates our present investigation of general physical processes that can cause the input states to become classical, and in particular, our search for associated overall input-output processes that optimally mimic quantum teleportation.

\section{General model of classical teleportation}

In our classical-teleportation model, first, we assume that the input particle with a state $\rho_{\rm{in},m}$ undergoes a generic physical process and decays into a classical system which possesses a pre-existing state $\lambda^{A}_{\xi}(a_1,a_2,a_3)$. Since the general tomographic characterization of experimental teleportation in Eq.~(\ref{exp}) involves three physical properties of input states, the possible measurement outcomes $a_1,a_2,a_3\in\{+1,-1\}$ for the three particle properties ${A}_1$, ${A}_2$ and ${A}_3$, respectively, are used to characterize the pre-existing states. See Fig.~\ref{QT}(c).

Second, the state $\lambda^{A}_{\xi}$ of the decayed particle corresponds to the output state of the particle on Bob's side, which is described by a density operator, $\rho^{B}_{\xi}$. Combining all possible measurement outcomes of the pre-existing states with Bob's corresponding particle states ${\rho^{B}_{\xi}}$, there exist the following eight possible states describing the particle pair:
\begin{eqnarray}\label{unsteerable}
&&[\lambda^{A}_1(+1,+1,+1);\rho^{B}_{1}],
\hspace{0.3cm}[\lambda^{A}_2(+1,+1,-1);\rho^{B}_{2}], \nonumber \\
&&[\lambda^{A}_3(+1,-1,+1);\rho^{B}_{3}],
\hspace{0.3cm}[\lambda^{A}_4(+1,-1,-1);\rho^{B}_{4}], \nonumber \\
&&[\lambda^{A}_5(-1,+1,+1);\rho^{B}_{5}],
\hspace{0.3cm}[\lambda^{A}_6(-1,+1,-1);\rho^{B}_{6}], \\
&&[\lambda^{A}_7(-1,-1,+1);\rho^{B}_{7}],
\hspace{0.3cm}[\lambda^{A}_8(-1,-1,-1);\rho^{B}_{8}]. \nonumber
\end{eqnarray}
When the particle pairs probabilistically decay into one of the above states with a probability distribution $P(\lambda^{A}_{\xi})$, the final state is a mixture of $[\lambda^{A}_{\xi};\rho^{B}_{\xi}]$ in general.

Finally, according to the overall input-output process described above, the four output states $\rho_{\rm{out},m}$ used for constructing the process matrix [Eq.~(\ref{exp})] are
\begin{eqnarray}\label{outputs}
& \rho_{0}\!\!=\!\!\sum_{\xi=1,3,5,7}\!2P({\lambda^{A}_\xi}){\rho^{B}_{\xi}},  \rho_{1}\!\!=\!\!\sum_{\xi=2,4,6,8}\!2P({\lambda^{A}_\xi}){\rho^{B}_{\xi}},\nonumber
\\
&\rho_{+}\!\!=\!\!\sum_{\xi=1,2,3,4}\!2P({\lambda^{A}_\xi}){\rho^{B}_{\xi}},  \rho_{R}\!\!=\!\!\sum_{\xi=1,2,5,6}\!2P({\lambda^{A}_\xi}){\rho^{B}_{\xi}}. \label{rhoQPT}
\end{eqnarray}
It is assumed that the properties $A_1$, $A_2$ and $A_3$ correspond to the observed Pauli-$X$, Pauli-$Y$ and Pauli-$Z$ matrices, respectively. From Eqs.~(\ref{exp}) and (\ref{rhoQPT}), classical teleportation is then tomographically characterized by
\begin{equation}
\chi_{\rm{CT}}=\frac{1}{2}\left[\begin{array}{cc}\rho_{0} & \rho_{+}+i\rho_{R}-\tilde{I}_{\rm{C}} \\ \rho_{+}-i\rho_{R}-\tilde{I}_{\rm{C}}^{\dag} & \rho_{1}\end{array}\right],\label{ct}
\end{equation}
where $\tilde{I}_{\rm{C}}=e^{i\pi/4}(\rho_{0}+\rho_{1})/\sqrt{2}$.

The definition of classical-teleportation model also has the following operational meaning which describes what Alice and Bob are allowed to do and use.
 After receiving the state $\rho_{\rm{in},m}$ from Victor, by which Alice can use any classical means to infer a state that she wants Bob to prepare for his particle: $\rho_{\rm{out},m}=\sum_\xi P(\xi|m)\rho^{B}_{\xi}$, where the index $\xi$ is used to label possible states $\rho^{B}_{\xi}$ of Bob's particle with preparation probabilities $P(\xi|m)$ and $\sum_\xi P(\xi|m)=1$. Here Alice needs to inform Bob about the $\rho_{\rm{out},m}$, and Bob is assumed to be capable of preparing these states. In particular, the process of deduction made by Alice is classical in the sense that, after deduction, the state $\rho_{\rm{in},m}$ becomes classical and independent of observation. This implies
that the preparation probability can be rephrased as: $P(\xi|m)=P({\lambda^{A}_\xi})P(m|{\lambda^{A}_\xi})/P(m)$, where $\lambda^{A}_\xi$ is the
pre-existing state described in Eq. (\ref{unsteerable}). Without loss of generality, assuming a normal distribution of input state, $P(m)=1/2$, we arrive at Eq. (\ref{rhoQPT}) for $\rho_{\rm{out},m}$ of classical
teleportation and the subsequent $\chi_{\rm{CT}}$ (\ref{ct}). See also Fig.~\ref{QT}(c).

The above general model of classical teleportation generalizes the concepts and methods for describing and implementing classical processes given in existing models, which specify a classical process by describing a single system's initial state and its evolution \cite{Hsieh17,Chen20}. Herein, we show that particle pairs with the state $[\lambda^{A}_{\xi};\rho^{B}_{\xi}]$ (\ref{unsteerable}) can be used to describe and implement a classical process [Figs.~\ref{QT}(c) and (d)]. As will be shown later, this important feature of the introduced model not only makes it possible to explore the extent to which teleportation can be performed with classical particle pairs, but also enables a strict examination of what state resources of particle pairs are required to achieve truly nonclassical teleportation.

\section{Genuine quantum teleportation and its identification and quantification\\}
Suppose that the process fidelity of experimental teleportation $\chi_{\rm{expt}}$ and ideal teleportation, $\chi_{\rm{I}}$, $F_{\rm{expt}}$, is used to evaluate the
performance of the experimental teleportation. If $\chi_{\rm{expt}}$ can surpass any classical teleportation $\chi_{\rm{CT}}$ by
\begin{equation}
F_{\rm{expt}}> F_{\rm{CT}}\equiv\max_{\chi_{\rm{CT}}} tr(\chi_{\rm{CT}}\chi_{\rm{I}}),\label{fidelity}
\end{equation}
then $\chi_{\rm{expt}}$ is qualified as \textit{genuine quantum teleportation}. In other words, the constituents of $\chi_{\rm{expt}}$, such as the measurement for BSM and the pair shared between Alice and Bob, are truly quantum-mechanical.

The ability of classical teleportation can be evaluated by performing the following mathematical maximization task via semi-definite programming (SDP) \cite{Yalmip04,SDPT34}: $\max_{\tilde{\chi}_{\rm{CT}}} tr(\tilde{\chi}_{\rm{CT}}\chi_{\rm{I}})$,
such that $\tilde{\chi}_{\rm{CT}}\geq 0$ and $tr(\tilde{\chi}_{\rm{CT}})=1$, where $\tilde{\chi}_{\rm{CT}}$ is an unnormalized process matrix, and the two constraints ensure that $\chi_{\rm{CT}}$ satisfies the definitions of both the process fidelity and the density operator. The best ability of classical teleportation to optimally mimic ideal teleportation is
\begin{equation}
F_{\rm{CT}}\sim0.683.\label{0.683}
\end{equation}
Through Eq.~(\ref{fsp}), the criterion for genuine quantum teleportation~(\ref{fidelity}) with Eq.~(\ref{0.683}) can be rephrased in terms of the average state fidelity, $\bar{F}_{\rm{expt},\rm{s}}$, as
\begin{equation}
\bar{F}_{\rm{expt},\rm{s}}>\frac{1}{3}(2F_{\rm{CT}}+1)\sim0.789.\label{0.789}
\end{equation}
Compared to the criterion for outperforming the classical measure-prepare strategy, i.e., $\bar{F}_{\rm{expt},\rm{s}}>2/3\sim0.667$ \cite{Massar95} (equivalently, $F_{\rm{expt}}>1/2$; see Eq.~(\ref{1/2})), a stricter criterion (\ref{0.789}) is required for identifying genuine quantum teleportation.

In addition to revealing genuine quantum teleportation with the process fidelity criterion (\ref{fidelity}), two additional methods can also be used to quantitatively describe the difference between genuine quantum teleportation and classical teleportation, $\chi_{\rm{CT}}$:\\

\noindent(a) Any $\chi_{\rm{expt}}$ can be represented as
\begin{equation}
\chi_{\rm{expt}}=\alpha\chi_{\rm{QT}}+(1-\alpha)\chi_{\rm{CT}},\label{alpha}
\end{equation}
where $\chi_{\rm{QT}}$ denotes a process that cannot be described at all by $\chi_{\rm{CT}}$, such as ideal teleportation $\chi_{\rm{I}}$, and the quantum composition, $\alpha$, describes the minimum amount of $\chi_{\rm{QT}}$ that can be found in $\chi_{\rm{expt}}$. See Fig.~\ref{QT}(d). The quantum composition $\alpha$ can be obtained by minimizing the following quantity via SDP: $\alpha=\min_{\tilde{\chi}_{\rm{CT}}} 1-tr(\tilde{\chi}_{\rm{CT}})$ such that $\tilde{\chi}_{\rm{CT}}\geq 0$ and $\chi_{\rm{expt}}-\tilde{\chi}_{\rm{CT}}\geq 0$, where the unnormalized process matrix $\tilde{\chi}_{\rm{CT}}$ possesses the property $tr(\tilde{\chi}_{\rm{CT}})=tr[(1-\alpha)\chi_{\rm{CT}}]=1-\alpha$. Note that the constraint $\chi_{\rm{expt}}-\tilde{\chi}_{\rm{CT}}\geq 0$ ensures that $\chi_{\rm{QT}}$ is positive semi-definite.\\

\noindent(b) An experimental process can be characterized by the minimum amount of noise process, called the quantum robustness and denoted by $\beta$, which must be added such that $\chi_{\rm{expt}}$ becomes classical, i.e.,
\begin{equation}
\frac{(\chi_{\rm{expt}}+\beta\chi')}{(1+\beta)}=\chi_{\rm{CT}},\label{beta}
\end{equation}
where $\chi'$ is the noise process. See Fig.~\ref{QT}(d). The quantum robustness, $\beta$, can be obtained by minimizing the following quantity via SDP: $\beta=\min_{\tilde{\chi}_{\rm{CT}}} tr(\tilde{\chi}_{\rm{CT}})-1$ such that $\tilde{\chi}_{\rm{CT}}\geq 0$, $tr(\tilde{\chi}_{\rm{CT}})\geq 1$ and $\tilde{\chi}_{\rm{CT}}-\chi_{\rm{expt}}\geq 0$, where $\tilde{\chi}_{\rm{CT}}=(1+\beta)\chi_{\rm{CT}}$ and $tr(\tilde{\chi}_{\rm{CT}})=1+\beta$. The constraints $tr(\tilde{\chi}_{\rm{CT}})\geq 1$ and $\tilde{\chi}_{\rm{CT}}-\chi_{\rm{expt}}\geq 0$ ensure that $\beta\geq0$ and $\chi'$ is positive semi-definite, respectively.\\

Both methods are helpful for distinguishing between truly quantum teleportation and classical mimicries. In particular, if an experiment shows that $\alpha=0$ and $\beta=0$, $\chi_{\rm{expt}}$ is identified as classical teleportation, $\chi_{\rm{CT}}$. On the other hand, if $\alpha,\beta>0$ and $F_{\rm{expt}}> 0.683$, then $\chi_{\rm{expt}}$ is genuine quantum teleportation close to $\chi_{\rm{I}}$. Notably, the values of $\alpha$ and $\beta$ determine the extent of quantumness. For example, when $\alpha=1$, $\beta\sim0.464$, and $F_{\rm{expt}}=1$, the process can be quantified as ideal teleportation.

\section{Measure-prepare processes as special cases of $\chi_{\text{CT}}$}

We will show that the general classical-teleportation model $\chi_{\rm{CT}}$ (\ref{ct}) can describe the measure-prepare mimicry model \cite{Massar95}, including the extended version introduced in Ref. \cite{Cavalcanti17}. Moreover, we will prove that our classical-teleportation model outperforms such mimicry and achieves a better simulation of ideal teleportation.\\

\noindent (a) Measure-prepare strategy. If there exists a cheat who utilizes the measure-prepare strategy in experiments where QPT is used to characterize experimental teleportation \cite{Pirandola15,Ma12,Bao12,Olmschenk09,Steffen13}, Alice's measurements, which are randomly chosen with regard to the input states: $\ket{0}$, $\ket{1}$, $\ket{+}$, and $\ket{R}$, cause Victor's input state $\ket{m}$ to decay to
\begin{eqnarray}
&&\rho_{m}=\frac{1}{3}(\sum_{{a_1}}p_{{a_1},m}\ket{a_1}\!\!\bra{a_1}+\sum_{{a_2}}p_{{a_2},m}\ket{a_2}\!\!\bra{a_2}\nonumber\\
&&\ \ \ \ \ \ \ \ \ \ \ \ +\sum_{{a_3}}p_{{a_3},m}\ket{a_3}\!\!\bra{a_3}),\label{rm2}
\end{eqnarray}
where $\ket{a_1}$, $\ket{a_2}$, and $\ket{a_3}$ denote the eigenstates for the Pauli-$X$, Pauli-$Y$, and Pauli-$Z$ matrices, respectively. Note that the Pauli matrices have the following spectral decompositions: $X=\ket{+}\!\!\bra{+}-\ket{-}\!\!\bra{-}$, $Y=\ket{R}\!\!\bra{R}-\ket{L}\!\!\bra{L}$ and $Z=\ket{0}\!\!\bra{0}-\ket{1}\!\!\bra{1}$, where $\ket{-}=(\ket{0}-\ket{1})/\sqrt{2}$ and $\ket{L}=(\ket{0}- i\ket{1})/\sqrt{2}$. Each constituent, $\sum_{{a_k}}p_{{a_k},m}\ket{a_k}\!\!\bra{a_k}$ is the average output state obtained from measurements of the physical property $A_{k}$ with a randomly chosen probability $1/3$. See~Eq.~(\ref{rm}). Using our classical-teleportation model, such output states can be fully described by Eq.~(\ref{outputs}), with $P(\lambda^{A}_{\xi})=1/8$ for $\xi=1,2,...,8$, and
\begin{eqnarray}\label{recipe}
& \rho_{1}^{B}=\frac{\ket{0}\!\bra{0}+\ket{+}\!\bra{+}+\ket{R}\!\bra{R}}{3},
 \rho_{2}^{B}=\frac{\ket{1}\!\bra{1}+\ket{+}\!\bra{+}+\ket{R}\!\bra{R}}{3},\nonumber \\
& \rho_{3}^{B}=\frac{\ket{0}\!\bra{0}+\ket{+}\!\bra{+}+\ket{L}\!\bra{L}}{3},
 \rho_{4}^{B}=\frac{\ket{1}\!\bra{1}+\ket{+}\!\bra{+}+\ket{L}\!\bra{L}}{3},\nonumber \\
& \rho_{5}^{B}=\frac{\ket{0}\!\bra{0}+\ket{-}\!\bra{-}+\ket{R}\!\bra{R}}{3},
 \rho_{6}^{B}=\frac{\ket{1}\!\bra{1}+\ket{-}\!\bra{-}+\ket{R}\!\bra{R}}{3},\nonumber \\
& \rho_{7}^{B}=\frac{\ket{0}\!\bra{0}+\ket{-}\!\bra{-}+\ket{L}\!\bra{L}}{3},
 \rho_{8}^{B}=\frac{\ket{1}\!\bra{1}+\ket{-}\!\bra{-}+\ket{L}\!\bra{L}}{3}.
\end{eqnarray}

Let us take $\ket{m}=\ket{0}$ chosen by Victor as the input state for illustration purposes. According to Eq.~(\ref{rm2}), the state after Alice's measurements decays to $\rho_{0}=(\ket{0}\!\!\bra{0}+I)/3$. This state is identical to the result derived from Eqs.~(\ref{outputs}) and~(\ref{recipe}) by
\begin{eqnarray}
\rho_{0}&=&\frac{1}{4}\sum_{\xi=1,3,5,7}{\rho^{B}_{\xi}}\nonumber\\
&=&\frac{1}{6}(2\ket{0}\!\!\bra{0}+
\ket{+}\!\!\bra{+}+
\ket{-}\!\!\bra{-}+
\ket{R}\!\!\bra{R}+
\ket{L}\!\!\bra{L})\nonumber\\
&=&\frac{1}{3}(\ket{0}\!\!\bra{0}+I).\nonumber
\end{eqnarray}
For the other input states, the corresponding output states, $\rho_{m}=(\ket{m}\!\!\bra{m}+I)/3$, can be described in the same manner by using the general model of classical teleportation. The measure-prepare process for teleportation, therefore, is a special case of classical teleportation $\chi_{\rm{CT}}$ (\ref{ct}) with the following process matrix
 \begin{equation}
\chi_{\rm{CT}\!\!,\rm{MP}}=\frac{1}{6}\left[\begin{array}{cc}\ket{0}\!\!\bra{0}+I &  \ket{+}\!\!\bra{+}\!+\!i\ket{R}\!\!\bra{R}\!-\!\hat{I}  \\ \ket{+}\!\!\bra{+}\!-\!i\ket{R}\!\!\bra{R}\!-\!\hat{I}^{\dag} & \ket{1}\!\!\bra{1}+I\end{array}\right].\label{ctm}
\end{equation}
In addition to this instance of classical teleportation, as shown below, all possible circumstances involving the measure-prepare strategy \cite{Cavalcanti17} can be explained using the classical-teleportation model, and the associated teleportation can be tomographically described by Eq.~(\ref{ct}).

Since $\chi_{\rm{CT}\!\!,\rm{MP}}$ is classical, we have $\alpha=0$ and $\beta=0$. In particular, the measure-prepare process, $\chi_{\rm{CT}\!\!,\rm{MP}}$, is not an optimal classical teleportation. That is, its process fidelity of $\chi_{\rm{I}}$ and $\chi_{\rm{CT},\rm{MP}}$ is smaller than the best similarity, $F_{\rm{CT}}$ (\ref{0.683}), that can be achieved by $\chi_{\rm{CT}}$, i.e.,
\begin{eqnarray}
F_{\rm{expt},\rm{MP}}&\equiv&tr(\chi_{\rm{I}}\chi_{\rm{CT},\rm{MP}})=0.5<0.683,
\end{eqnarray}
where the process fidelity, $F_{\rm{expt},\rm{MP}}$, is consistent with the existing threshold obtained by the measure-prepare method~\cite{Massar95} (see also Eq.~(\ref{1/2})). It is possible that, while a demonstrated teleportation by entangled pairs has a process fidelity $F_{\rm{expt}}$ with the quality: $0.5<F_{\rm{expt}}\leq0.683$, and thus surpasses the measure-prepare mimicry, it still can be fully described by the general classical teleportation, $\chi_{\rm{CT}}$ (\ref{ct}). (See the example provided in Fig.~\ref{example}.)\\

\noindent (b) Measure-prepare model extended by Cavalcanti \textit{et al.} \cite{Cavalcanti17}. Compared to the ordinary measure-prepare model where there are no pairs shared between Alice and Bob, in the extended measure-prepare model \cite{Cavalcanti17}, Alice and Bob share particle pairs in separable states, $\rho^{AB}=\sum_{\kappa}p_{\kappa}\rho^{A}_{\kappa}\otimes\rho^{B}_{\kappa}$. Moreover, in extending the direct measurements to the unknown states $\rho_{\rm{in}, m}$ in the ordinary measure-prepare strategy, Alice first measures the input states $\rho_{\rm{in}, m}$ from Victor and her share of the pair with the POVM elements $M^{VA}_a$ and obtains the outcome $a$ with a probability $P(a|m)$. The unnormalized teleported state at Bob's side is then given as
\begin{eqnarray}
\tilde{\rho}_{\rm{out},m,a}&=&tr_{VA}[(M^{VA}_a\otimes I^{B})(\rho_{\rm{in},m}\otimes \rho^{AB})]\nonumber \\ \label{unnteleported_state}
&=&tr_{V}[M^{VB}_a(\rho_{\rm{in},m}\otimes I^{B})],
\end{eqnarray}
where the operator $M^{VB}_a$ is defined as
\begin{eqnarray}
M^{VB}_a&=&tr_{A}[(M^{VA}_a\otimes I^{B})(I^{V}\otimes \rho^{AB})]\nonumber \\
&=&\sum_{\kappa}p_{\kappa}tr_{A}[(M^{VA}_a\otimes I^{B})(I^{V}\otimes \rho^{A}_{\kappa}\otimes\rho^{B}_{\kappa})]\nonumber \\
&=&\sum_{\kappa}p_{\kappa}M^{V}_{a|\kappa}\otimes\rho^{B}_{\mathcal{\kappa}}, \label{channel_operator}
\end{eqnarray}
and $M^{V}_{a|\kappa}=tr_{A}[M^{VA}_a(I^{V}\otimes\rho^{A}_{\kappa})]$; see Eq.~(5) in Ref.~\cite{Cavalcanti17} for more details.
The probability $P(a|m)$ is the normalization factor $tr(\tilde{\rho}_{\rm{out},m,a})$, i.e., $P(a|m)=tr(\tilde{\rho}_{\rm{out},m,a})$,
and the normalized teleported state is
\begin{eqnarray}
\rho_{\rm{out},m,a}=\frac{\tilde{\rho}_{\rm{out},m,a}}{P(a|m)}.\label{teleported_state}
\end{eqnarray}
After Bob receives the measurement outcomes from Alice, he corrects the states $\rho_{\rm{out},m,a}$ with the unitary operator $U_a$ for each measurement outcome $a$ and obtains the output states
\begin{eqnarray}
\rho_{\rm{out},m}=\sum_{a}P(a|m)U_a\rho_{\rm{out},m,a}U_a^\dagger. \label{out_m}
\end{eqnarray}

To show that the teleported state in Eq.~(\ref{out_m}) can be described by our classical-teleportation model, let us consider the same examples of Alice's measurements $M^{VA}_a$ as those discussed by Cavalcanti \textit{et al.} \cite{Cavalcanti17,Supic19}:\\

\noindent (i) Product measurement. The POVM elements $M^{VA}_a$ in the form of product measurement are {$M^{VA}_a=M^{V}_a\otimes M^{A}_a$}, where $M^{V}_a$ and $M^{A}_a$ are measurement operators acting on the state to be teleported and Alice's half of the shared pair with the measurement outcome $a$. The outcome states in Eq.~(\ref{out_m}) at Bob's side become
\begin{eqnarray}\label{rho_out_m_product}
\rho_{\rm{out},m}=\sum_{a}P(a|m)U_a\rho^{B}_{a}U_a^\dagger,
\end{eqnarray}
where $P(a|m)=tr(M^{V}_a\rho_{\rm{in},m})$ and
\begin{eqnarray}\label{rho_out_m_a_product}
\rho^{B}_{a}=\frac{\sum_{\kappa}p_{\kappa}tr( M^{A}_a\rho^{A}_{\kappa})\rho^{B}_{\kappa}}{\sum_{\kappa}p_{\kappa}tr( M^{A}_a\rho^{A}_{\kappa})}.
\end{eqnarray}
The output states in (\ref{rho_out_m_product}) can be described by our classical-teleportation model in (\ref{unsteerable}) and (\ref{outputs}) with \begin{eqnarray}\label{p_xi_solution}
P(\lambda^{A}_{\xi})=\frac{1}{8}
\end{eqnarray}
for $\xi=1,2,...,8$, and Bob's particle states $\rho_{\xi}^{B}$,
\begin{eqnarray}\label{recipe1}
& \rho_{1}^{B}=\sum_{a}[P(a|0)+P(a|+)+P(a|R)-2P(a)]U_a\rho^{B}_{a}U_a^\dagger,\nonumber \\
& \rho_{2}^{B}=\sum_{a}[P(a|1)+P(a|+)+P(a|R)-2P(a)]U_a\rho^{B}_{a}U_a^\dagger,\nonumber \\
& \rho_{3}^{B}=\sum_{a}[P(a|0)+P(a|+)+P(a|L)-2P(a)]U_a\rho^{B}_{a}U_a^\dagger,\nonumber \\
& \rho_{4}^{B}=\sum_{a}[P(a|1)+P(a|+)+P(a|L)-2P(a)]U_a\rho^{B}_{a}U_a^\dagger,\nonumber \\
& \rho_{5}^{B}=\sum_{a}[P(a|0)+P(a|-)+P(a|R)-2P(a)]U_a\rho^{B}_{a}U_a^\dagger,\nonumber \\
& \rho_{6}^{B}=\sum_{a}[P(a|1)+P(a|-)+P(a|R)-2P(a)]U_a\rho^{B}_{a}U_a^\dagger,\nonumber \\
& \rho_{7}^{B}=\sum_{a}[P(a|0)+P(a|-)+P(a|L)-2P(a)]U_a\rho^{B}_{a}U_a^\dagger,\nonumber \\
& \rho_{8}^{B}=\sum_{a}[P(a|1)+P(a|-)+P(a|L)-2P(a)]U_a\rho^{B}_{a}U_a^\dagger,\nonumber\\
\end{eqnarray}
where $P(a)=(P(a|0)+P(a|1))/2=(P(a|+)+P(a|-))/2=(P(a|R)+P(a|L))/2$.

Let us take $\ket{m}=\ket{+}$ for illustration purposes. From Eqs.~(\ref{outputs}) and~(\ref{recipe1}), the output state is
\begin{eqnarray}
\rho_{\rm{out},+}&=&\frac{1}{4}\sum_{\xi=1,2,3,4}{\rho^{B}_{\xi}}\nonumber\\
&=&\sum_{a}[\frac{1}{2}(P(a|0)+P(a|1))+P(a|+)\nonumber\\
&& \ \ \ \ \ +\frac{1}{2}(P(a|R)+P(a|L))-2P(a)]U_a\rho^{B}_{a}U_a^\dagger\nonumber\\
&=&\sum_{a}P(a|+)U_a\rho^{B}_{a}U_a^\dagger,\nonumber
\end{eqnarray}
which shows that the output states $\rho_{\rm{out},+}$ in (\ref{rho_out_m_product}) can be described by $P(\lambda^{A}_{\xi})$ and $\rho_{\xi}^{B}$ in (\ref{recipe1}).
For the other input states, the corresponding output states, $\rho_{\rm{out},m}$ (\ref{rho_out_m_product}), can be described in the same manner by using our classical-teleportation model. (See Appendix C for a complete proof.)\\

\noindent (ii) Bell-state measurement. For a partial Bell-state measurement with measurement operators $M^{VA}_1=\ket{\phi^+}\!\!\bra{\phi^+}$ and $M^{VA}_2=I-\ket{\phi^+}\!\!\bra{\phi^+}$, the corresponding $M^{V}_{a|\kappa}$ in Eq.~(\ref{channel_operator}) can be represented by the transpose of $\rho^{A}_{\kappa}$, i.e., $M^{V}_{1|\kappa}={\rho^{AT}_{\kappa}}$ and $M^{V}_{2|\kappa}=I-{\rho^{AT}_{\kappa}}$. We then have the following operators $M^{VB}_a$~(\ref{channel_operator}): $M^{VB}_1=\sum_{\kappa}p_{\kappa}{\rho^{AT}_{\kappa}}\otimes\rho^{B}_{\kappa}$ and $M^{VB}_2=\sum_{\kappa}p_{\kappa}(I-{\rho^{AT}_{\kappa}})\otimes\rho^{B}_{\kappa}$.

As the measurement outcome is $a=a'$, through Eqs.~(\ref{unnteleported_state}) and (\ref{teleported_state}), the output states $\rho_{\rm{out},m,a'}$ of the input states $\rho_{\rm{in},m}=\ket{m}\!\!\bra{m}$ are
\begin{eqnarray}
\rho_{\rm{out},m,a'}&=&\frac{1}{P(m|a')}\sum_{\kappa}p_{\kappa}tr(M^{V}_{a'|\kappa}\ket{m}\!\!\bra{m})\rho^{B}_{\kappa}\nonumber \\
&=&\frac{1}{P(m|a')}\sum_{\kappa}p_{\kappa}P(m|\kappa,a')\rho^{B}_{\kappa},\label{noroutput}
\end{eqnarray}
where $P(m|a')=\sum_{\kappa}p_{\kappa}P(m|\kappa,a')$. We have $P(m|\kappa,1)=tr({\rho^{AT}_{\kappa}}\ket{m}\!\!\bra{m})$ for $a'=1$ and $P(m|\kappa,2)=1-tr({\rho^{AT}_{\kappa}}\ket{m}\!\!\bra{m})$ for $a'=2$. The output states (\ref{noroutput}) conditioned on $a'$ can be described by our classical model, (\ref{unsteerable}) and (\ref{outputs}), under the distribution
\begin{eqnarray}\label{P_lambda_A_xi_a}
P(\lambda^{A}_{\xi}|a')=\sum_{\kappa}P(\lambda^{A}_{\xi},\kappa|a')
\end{eqnarray}
and
\begin{eqnarray}\label{rho_B_xi_a}
\rho^{B}_{\xi,a'}=\frac{1}{P(\lambda^{A}_{\xi}|a')}\sum_{\kappa}p_{\kappa}P(\lambda^{A}_{\xi}|\kappa,a')\rho^{B}_{\kappa},
\end{eqnarray}
where $P(\lambda^{A}_{\xi}|a')$ and $\rho^{B}_{\xi,a'}$ denote the pre-existing recipe of classical teleportation used in case $a'$; see Appendix C.
Take $\ket{m}=\ket{+}$ for illustration purposes. By substituting Eqs.~(\ref{P_lambda_A_xi_a}) and (\ref{rho_B_xi_a}) into Eq.~(\ref{outputs}), we have the same result of Eq.~(\ref{noroutput}) by
\begin{eqnarray}
\rho_{\rm{out},+,a'}&=&\sum_{\xi=1,2,3,4}2P(\lambda^{A}_{\xi}|a'){\rho^{B}_{\xi,a'}}\nonumber \\
&=&\sum_{\xi=1,2,3,4}2\sum_{\kappa}p_{\kappa}P(\lambda^{A}_{\xi}|\kappa,a')\rho^{B}_{\kappa}\nonumber \\
&=&2\sum_{\xi=1}^{8}\sum_{\kappa}p_{\kappa}P(\lambda^{A}_{\xi}|\kappa,a')P(+|\lambda^{A}_{\xi},\kappa,a')\rho^{B}_{\kappa}\nonumber \\
&=&\frac{1}{P(m|a')}\sum_{\kappa}p_{\kappa}P(+|\kappa,a')\rho^{B}_{\kappa},\nonumber
\end{eqnarray}
under the assumption of a uniform distribution of the input states for QPT, i.e., $P(m|a')=1/2$.

When taking the correction $U_a$ for each case $a$ in Eq.~({\ref{out_m}}) into account, the output states $\rho_{\rm{out},m}$, which are the linear combination of $U_a\rho_{\rm{out},m,a}U_a^\dagger$, become
\begin{eqnarray}\label{rho_out_m_U_a}
\rho_{\rm{out},m}&=&\sum_{a}P(a|m)U_a\left(\sum_{\xi|m}2P(\lambda^{A}_{\xi}|a)\rho^{B}_{\xi,a}\right)U_a^\dagger\nonumber \\
&=&\sum_{\xi|m}\sum_{a}2P(\lambda^{A}_{\xi},a)U_a\rho^{B}_{\xi,a}U_a^\dagger,
\end{eqnarray}
where $\xi|m$ denotes the pre-existing states $\lambda^{A}_{\xi}$ that possess a certain measurement result $m$, e.g., $\xi|+=1,2,3,4$ indicates that the states $\lambda^{A}_{1}$, $\lambda^{A}_{2}$, $\lambda^{A}_{3}$, and $\lambda^{A}_{4}$ have the measurement outcome $a_{1}=1$, i.e., $m=+$. (see Eqs.~(\ref{unsteerable}) and (\ref{outputs}).)
The output states $\rho_{\rm{out},m}$ can thus be expressed in the form of (\ref{outputs}) with
\begin{eqnarray}\label{P_lambda_A_xi_m}
P({\lambda}^{A}_{\xi})=\sum_{a}P(\lambda^{A}_{\xi},a),
\end{eqnarray}
and
\begin{eqnarray}\label{rho_B_xi_m}
{\rho}^{B}_{\xi}=\sum_{a}P(a|\lambda^{A}_{\xi})U_a\rho^{B}_{\xi,a}U_a^\dagger.
\end{eqnarray}
(See Appendix C for a more detailed explanation.)\\

From examples (i) and (ii), we conclude that the extended measure-prepare model proposed in Ref.~\cite{Cavalcanti17}, where separable states are used as the source, can always be described by our classical-teleportation model. That is, Alice's measurements on the unknown input state and her particle of the shared pair in a separable state make the state of Bob's particle classical as though the teleportation output were composed of pre-existing states. Since processes that can be described by such a model are classical, we have $\alpha=0$ and $\beta=0$. In particular, the best capability of the extended measure-prepare model to mimic ideal teleportation in terms of the process fidelity and average state fidelity are $F_{\rm{expt}}=1/2$ and $\bar{F}_{\rm{expt},\rm{s}}=2/3$ \cite{Cavalcanti17}, as obtained via SDP that conditioned on separable states as shared pairs. Our classical-teleportation model outperforms such mimicry and can achieve a better simulation of ideal teleportation, as shown in Eqs.~(\ref{0.683}) and~(\ref{0.789}).

\section{State resources required for genuine quantum teleportation}

To examine the state resources of particle pairs required to enable genuine quantum teleportation, we assume that the EPR pair used for teleportation [Fig.~\ref{QT}(a)] is replaced by a pair with an unknown experimental state, denoted as $\rho_{\rm{expt}}$. See Fig.~\ref{QT}(d). Moreover, we suppose that all the other elements required for the teleportation process are ideal. In the following, we use the process fidelity criterion~(\ref{fidelity}) to show that, as $\rho_{\rm{expt}}$ possesses EPR steerability, the resulting teleportation outperforms general classical teleportation.

\begin{figure}[t]
\centering
\includegraphics[width=7cm]{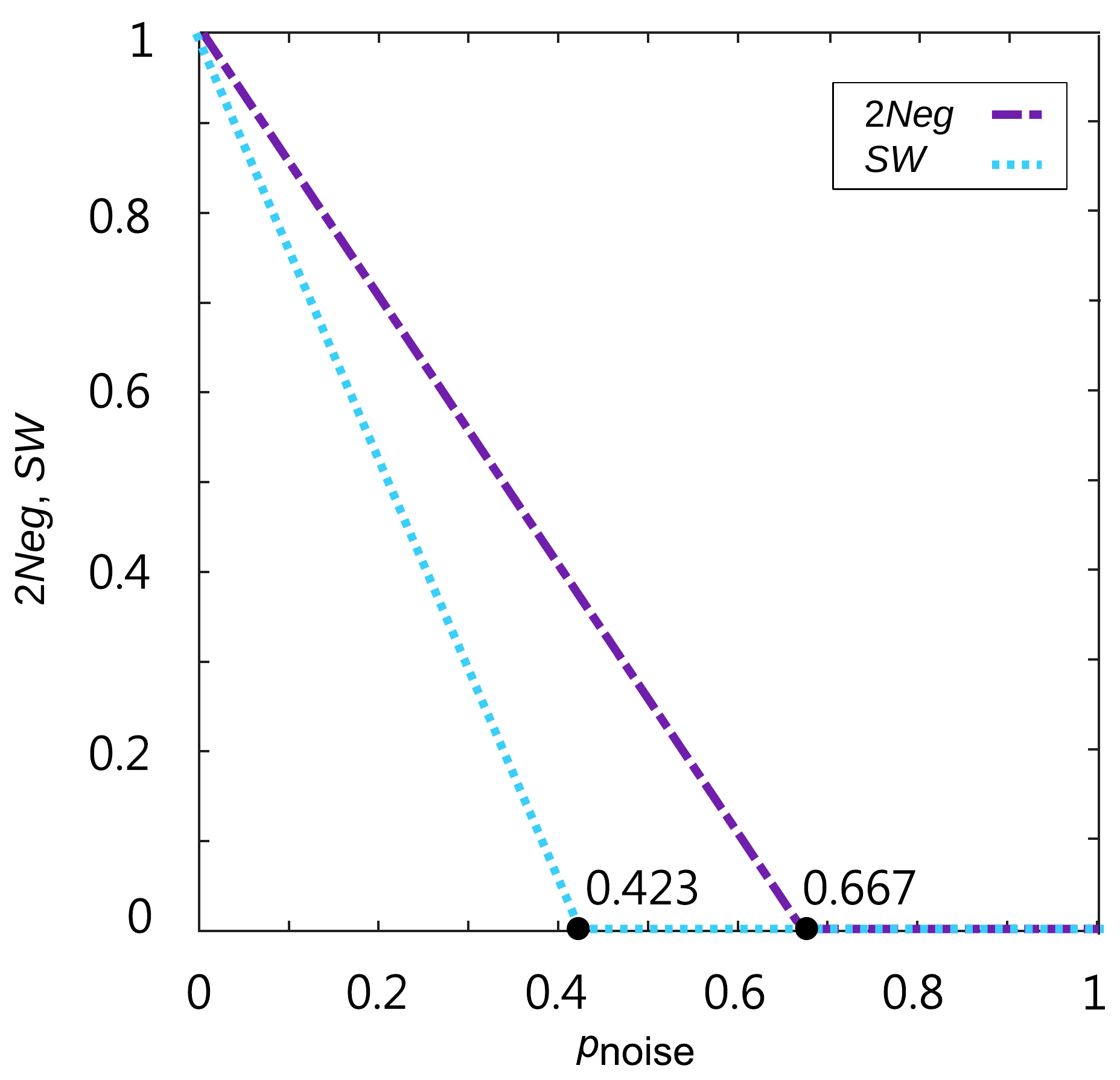}
\caption{Entanglement and steerability of $\rho_\text{{W}}(p_{\text{noise}})$. We use the negativity, $Neg(\rho_\text{{W}})\equiv(tr\sqrt{({\rho_\text{{W}}^{T_A}})^\dagger \rho_\text{{W}}^{T_A}}-1)/2$ \cite{vidal2002computable}, to measure the entanglement of $\rho_\text{{W}}$, where $\rho_\text{{W}}^{T_A}$ denotes the partial transpose of $\rho_\text{{W}}$. In addition, we use the steerable weight, $SW(\rho_\text{{W}})$~\cite{skrzypczyk2014quantifying}, to quantify the steerability of $\rho_\text{{W}}$. A steering experiment can be completely characterized by an ``assemblage'' $\{\rho_{a_k|k}\}_{a_k,k}$, i.e., the set of unnormalized states which Alice steers Bob into, where $k$ denotes Alice's choice of measurement setting {$A_k$} and $a_k$ represents the measurement outcome. Here, we consider the Pauli matrices, $X$, $Y$ and $Z$, as the observables for Alice's measurements performed on her half of the qubit pair of $\rho_\text{{W}}$. The assemblage $\rho_{a_k|k}$ can be decomposed as $\rho_{a_k|k}=\mu\rho_{a_k|k}^{\rm{US}}+(1-\mu)\rho_{a_k|k}^{\rm{S}}$ $ \forall$ $a_k,k $, where $\{\rho_{a_k|k}^{\rm{US}}\}_{a_k,k}$ ($\{\rho_{a_k|k}^{\rm{S}}\}_{a_k,k}$) denotes the unsteerable (steerable) assemblage and $0\leq\mu\leq1$. $SW(\rho_\text{{W}})$ is defined as $SW(\rho_\text{{W}})=1-\mu^*$, where $\mu^*$ denotes the maximum $\mu$ and can be obtained from SDP; see~\cite{skrzypczyk2014quantifying} for more details. That is, $SW(\rho_\text{{W}})$ quantifies the minimum amount of steerable resource needed to reproduce the given assemblage $\{\rho_{a_k|k}\}_{a_k,k}$.}
\label{qcorrelations}
\end{figure}

\begin{figure*}[t]
\centering
\includegraphics[width=18.54cm]{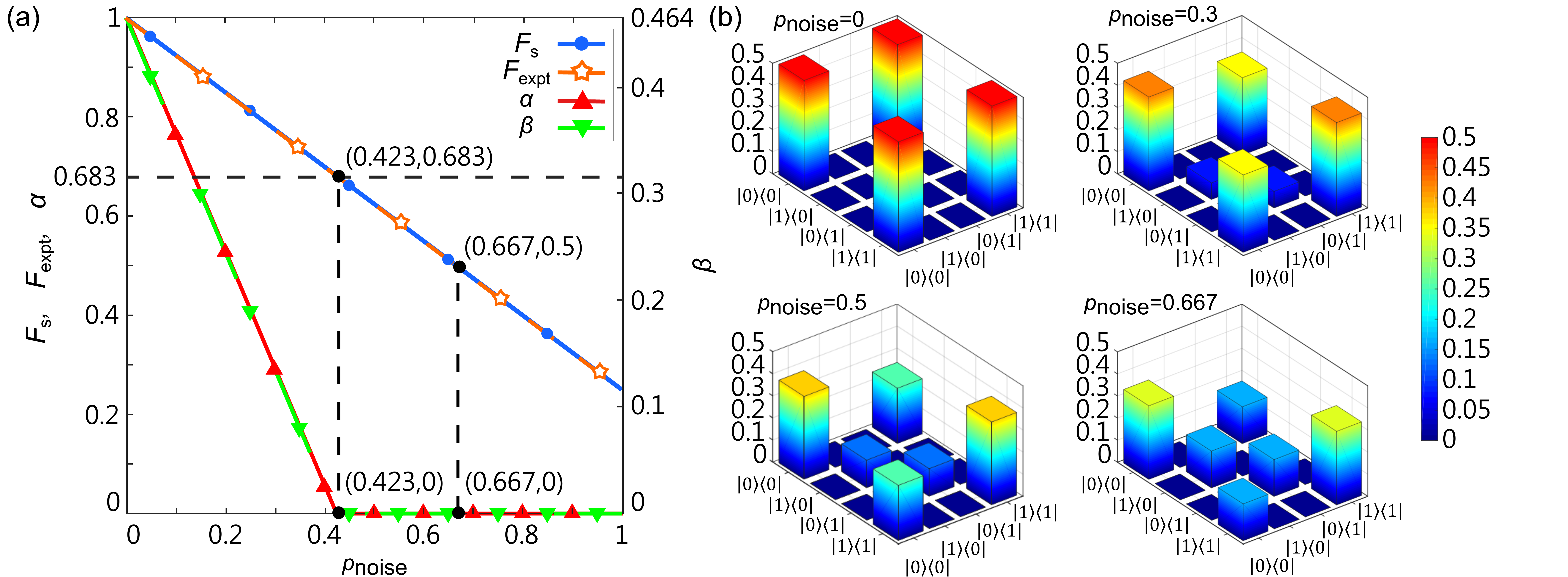}
\caption{Identifying genuine quantum teleportation under a noisy entanglement source. In particular, an experimental teleportation $\chi_{\rm{expt}}(p_{\rm{noise}})$ is affected by contaminated EPR pairs $\rho_{\rm{{W}}}(p_{\rm{noise}})$ (\ref{rw}). (a) The degraded teleportation can be evaluated using the process fidelity criterion (\ref{fidelity}), $\alpha$ (\ref{alpha}) and $\beta$ (\ref{beta}). (b) Tomographic details of the noisy teleportation processes. The process matrices of teleportation under four different noise intensities in (a) are demonstrated. Here, $\chi_{\rm{expt}}(0)=\chi_{\rm{I}}$ (see Eq.~(\ref{i})), and $\chi_{\rm{expt}}(0.667)$ has the same form as $\chi_{\rm{CT},\rm{MP}}$ (\ref{ctm}). The manner in which $\chi_{\rm{expt}}(0.5)$ can be represented as a $\chi_{\rm{CT}}$ is shown in Appendix E.}
\label{example}
\end{figure*}

First, if the state $\rho_{\rm{expt}}$ is used for teleportation under the assumptions given above, the corresponding process fidelity, $F_{\rm{expt}}$, will be equal to the state fidelity of $\rho_{\rm{expt}}$ and the ideal state of the EPR pair, defined by the equation $F_{\rm{{s}}}\equiv tr(\rho_{\rm{expt}}\!\ket{\phi^{+}}\!\!\bra{\phi^{+}})$. (See Appendix D for associated derivation.) This implies that the process fidelity criterion~(\ref{fidelity}) can be rephrased as
\begin{equation}
F_{\rm{{s}}}=F_{\rm{expt}}> F_{\rm{CT}}\sim0.683.\label{sfidelity}
\end{equation}
In other words, if a particle pair possesses a state which satisfies the above criterion, then Alice and Bob can perform genuine quantum teleportation.

Second, since the pre-existing state $\lambda^{A}_{\xi}$ of Alice's particle determines the specific output state of Bob's particle, $\rho^{B}_{\xi}$, the description of the particle-pair state related to Eq.~(\ref{unsteerable}) can be understood by a local hidden state model \cite{Wiseman07,skrzypczyk2014quantifying, piani2015necessary, gallego2015resource, uola2019quantum}, where $\rho^{B}_{\xi}$ denotes a local hidden state (LHS). In general, a particle pair by which Alice can control Bob's particle to invalidate the prediction of the LHS model is said to possess EPR steerability. By contrast, the LHS model constrains particle-pair states to be unsteerable.

Third, the criterion for the state fidelity $F_{\rm{{s}}}$ (\ref{sfidelity}) can also be utilized to certify EPR steering. As shown in Ref.~\cite{Lu20}, the maximum state fidelity between $\rho_{\rm{expt}}$ and $\ket{\phi^+}$ that can be achieved using unsteerable states under the LHS model (\ref{unsteerable}) is $F_{\rm{CT}}$. In other words, satisfying criterion (\ref{sfidelity}) indicates that $\rho_{\rm{expt}}$ has EPR steerability. The classical threshold $F_{\rm{CT}}$ can be beaten if and only if the shared resource state is steerable under the criterion (\ref{sfidelity}) which certifies both EPR steering and genuine quantum teleportations. Therefore, EPR steering empowers genuine quantum teleportations. Compared to the process fidelity criterion introduced in Ref.~\cite{Hsieh17}, i.e., $F_{\rm{expt}}>0.683$, which identifies whether an experimental teleportation can go beyond the simulation of classical evolution, the present model strictly examines the state resources required to achieve genuine quantum teleportation.

The maximum state fidelity $F_{\rm{{s}}}$ that can be attained using separable states is only $0.5$. For particle pairs with $0.5<F_{\rm{{s}}}\leq0.683$, the pairs are entangled, but their resulting teleportations do not surpass the general classical teleportation. In other words, not all entangled states can facilitate genuine quantum teleportation. For example, as will be illustrated later (see Figs.~\ref{qcorrelations} and \ref{example} and the related discussions), entangled but unsteerable states can be used to perform nonclassical teleportation with respect to the measure-prepare method, but are ineffective in achieving genuine quantum teleportation.

\begin{table*}
\caption{\label{table1}\label{table1}Identifying genuine quantum teleportation in practical experiments. With the process fidelity criterion given in Eq.~(\ref{fidelity}), we illustrate herein the examination outcomes of experimental teleportations performed with five different quantum teleportation technologies. For the case where the process fidelity, which is derived from either active or passive teleportation experiments, satisfies the process criterion, the experimental process is identified as genuine quantum teleportation (GQT) and is marked with a $\surd$. For experiments that can perform the measurements for QPT and obtain the process matrix $\chi_{\rm{expt}}$, the quantum composition $\alpha$ and quantum robustness $\beta$ are both measurable ($\surd$) in principle to quantify the degree of quantumness in $\chi_{\rm{expt}}$.}
\begin{ruledtabular}
\begin{tabular}{@{}llllllll}
 Quantum technology & & Process fidelity &   & GQT&$\alpha$, $\beta$ \\

 & &  Active & Passive & &measurable? \\
 \hline
 Photonic qubits & Polarization~\cite{Ma12, Ren17} & $75\% ^{*}$~\cite{Ma12}    & $70\% ^{*}$~\cite{Ren17}  &$\surd$&  \\
 &Time-bins~\cite{sun2016quantum} &77\%& 84\% &$\surd$&$\surd$ \\
NMR~\cite{Nielsen98}    & &$\sim90\%$&&$\surd$&$\surd$   \\

Atomic ensembles &Cold discrete-variable matter-to-matter~\cite{Bao12}& &87\%&$\surd$&$\surd$ \\

Trapped atoms& Trapped ions~\cite{riebe2007quantum}&73\% & &$\surd$&$\surd$ \\

                 &Trapped ions and photonic carriers~\cite{Olmschenk09} & &84\% &$\surd$&$\surd$ \\

                 &Neutral atoms in an optical cavity~\cite{nolleke2013efficient}& &$82\% ^{*}$&$\surd$ \\

Solid state &Polarization qubit to rare-earth crystal~\cite{bussieres2014quantum}& &$83.5\% ^{*}$ &$\surd$&\\

                 &Superconducting qubits on chip~\cite{Steffen13}&53.9\%&65.5\%&&$\surd$ \\

                 &Nitrogen-vacancy centres in diamonds~\cite{Pfaff14}&$65.5\% ^{*}$&&& \\
\end{tabular}
\end{ruledtabular}
\begin{flushleft}\footnotesize{$^{*}$We use the reported experimental values averaged over possible state fidelities for a finite number of input and output states to represent the experimental average state fidelity $\bar{F}_{\text{expt},\text{s}}$. The process fidelity $F_{\text{expt}}$ is then derived from $\bar{F}_{\text{expt},\text{s}}$ through Eq.~(\ref{fsp}) \cite{Gilchrist05}.}\end{flushleft}
\end{table*}

\section{Examples of identifying genuine quantum teleportation}

\subsection{Teleportation implemented with contaminated EPR pairs}
\noindent Any noise process due to environmental interactions or imperfections which arise in actual implementations can decrease the amount of genuine quantum teleportation present in an experimental teleportation. At worst, the experimental results can be fully explained using the classical-teleportation model; in particular, $\chi_{\rm{expt}}$ is a process matrix of classical teleportation, $\chi_{\rm{CT}}$ (\ref{ct}) [Fig.~\ref{QT}(d)].

Here, we assume that the EPR pair $\ket{\phi^+}$ used for teleportation [Fig.~\ref{QT}(a)] is contaminated by white noise and becomes
\begin{equation}
\rho_{\rm{{W}}}(p_{\rm{noise}})=(1-p_{\rm{noise}})\ket{\phi^{+}}\!\!\bra{\phi^{+}}+\frac{p_{\rm{noise}}}{4}I\otimes I,\label{rw}
\end{equation}
where $p_{\rm{noise}}$ denotes the intensity of the white noise. Figure~\ref{qcorrelations} illustrates the change in the entanglement and EPR steerability of $\rho_{\rm{{W}}}$ with the noise intensity. Note that the entanglement is measured by the negativity \cite{vidal2002computable}, $Neg$, and the steerability is quantified by the steerable weight \cite{skrzypczyk2014quantifying}, $SW$. As shown, $\rho_{\rm{{W}}}(p_{\rm{noise}})$ is identified as entangled for $0\leq p_{\rm{noise}}<0.667$ and as a steerable state for $0\leq p_{\rm{noise}}<0.423$. Notably, $\rho_{\rm{{W}}}(p_{\rm{noise}})$ is entangled but unsteerable for $0.423\leq p_{\rm{noise}}<0.667$.

To examine the effect of a noisy entanglement source, we suppose that all the other operations for teleportation are ideal, e.g., BSM. (See the scenario considered for Eq.~(\ref{sfidelity}) where $\rho_{\rm{expt}}=\rho_{\rm{{W}}}(p_{\rm{noise}})$.) Genuine quantum teleportation present in experimental processes under these conditions, denoted by $\chi_{\rm{expt}}(p_{\rm{noise}})$, can be characterized quantitatively by the fidelity criterion~(\ref{fidelity}), $\alpha$ (\ref{alpha}) and $\beta$ (\ref{beta}). See Fig.~\ref{example}. For $0\leq p_{\rm{noise}}<0.423$, $\chi_{\rm{expt}}(p_{\rm{noise}})$ is identified as genuine quantum teleportation, under the steerable state, $\rho_{\rm{{W}}}(p_{\rm{noise}})$ (see Fig.~\ref{qcorrelations}).

It is worth noting that, when the noise intensity, $p_{\rm{noise}}$, lies in the range $0.423 \leq p_{\rm{noise}} < 0.667$, i.e., $0.5 < F_{\rm{s}}\leq 0.683$, the corresponding state of the shared pair $\rho_{\rm{{W}}}(p_{\rm{noise}})$ is identified as entangled but unsteerable. Moreover, for $0.5 < F_{\rm{expt}}\leq 0.683$ (i.e., $0.667 <\bar{F}_{\rm{expt},\rm{s}}\leq 0.789$), the resulting teleportation process cannot be simulated by the measure-prepare strategy, but can still be described by the classical-teleportation model, $\chi_{\rm{CT}}$, i.e., $\alpha=0$ and $\beta=0$. (See Appendix E for a concrete example with $p_{\rm{noise}}=0.5$.) When $p_{\rm{noise}} \geq 0.667$, $\chi_{\rm{expt}}(p_{\rm{noise}})$ are classical with respect to the measure-prepare strategy.

\subsection{Genuine quantum teleportation in practical experiments}
\noindent In order to illustrate that our formalism for characterizing and identifying genuine quantum teleportation is experimentally feasible, we review herein the measurements performed in several existing teleportation experiments and show that the methods introduced in Eqs.~(\ref{fidelity}), (\ref{alpha}) and (\ref{beta}) can be used to evaluate the experimental results.

If a teleportation experiment is capable of implementing measurements of QPT, the experimental process matrix $\chi_{\rm{expt}}$ derived from the QPT measurements can be readily quantified in terms of $\alpha$ (\ref{alpha}) and $\beta$ (\ref{beta}). Moreover, its process fidelity can be examined using the process fidelity criterion~(\ref{fidelity}) or state fidelity criterion~(\ref{0.789}). To implement QPT, the capability of preparing four input states of teleportation, namely $\ket{0}$, $\ket{1}$, $\ket{+}$, and $\ket{R}$, is required. Moreover, it is necessary to tomographically analyze their corresponding output states, from which $\chi_{\rm{expt}}$ can be obtained according to Eq.~(\ref{exp}). Once $\chi_{\rm{expt}}$ has been measured, the process fidelity $F_{\rm{expt}}$ can be determined and $\alpha$ and $\beta$ can be calculated using SDP. It is worth noting that the experimental average state fidelity, $\bar{F}_{\rm{expt},\rm{s}}$, can also be obtained by measuring $F_{\rm{expt}}$ via Eq.~(\ref{fsp})~\cite{Gilchrist05}. Notably, our methods for evaluating $\chi_{\rm{expt}}$ can be applied to experiments where either Bob can apply conditional local operations in real-time before Victor's verification (i.e., active teleportation), or feed-forward is not realized or is simulated in post-processing (i.e., passive teleportation) \cite{Pirandola15}.

In the five quantum teleportation technologies reported in \cite{Nielsen98,Ma12, Ren17, Bao12, Olmschenk09, Steffen13, Pfaff14, sun2016quantum,riebe2007quantum,nolleke2013efficient, bussieres2014quantum}, the experimental teleportations are either examined by implementing tomographic experiments to obtain full knowledge of $\chi_{\rm{expt}}$ and the process fidelity $F_{\rm{expt}}$, or evaluated by measuring the state fidelity averaged over the possible fidelities of the input states $\rho_{\rm{in},m}$ and output states $\rho_{\rm{out},m}$ for a finite number of input and output states. Table~\ref{table1} summarizes the experimental fidelities in these experiments and indicates whether the corresponding results can be identified as genuine quantum teleportation in accordance with the fidelity criterion~(\ref{fidelity}).

\section{Summary and outlook}

We have developed herein a new benchmark consisting of fidelity criteria and quantitative identifications to examine the extent to which quantum teleportation can ultimately be performed without the use of maximally entangled EPR pairs. A general model of classical teleportation has additionally been introduced to evaluate the prescribed quantum teleportation of an experimental process. We have shown that classical teleportation can fully describe the measure-prepare procedure for teleportation \cite{Massar95}, including the extended version proposed in Ref. \cite{Cavalcanti17}. In particular, a new state fidelity criterion of $\bar{F}_{{\rm{expt}},{\rm{s}}}>0.789$, i.e., the process fidelity of $F_{\rm{expt}}>0.683$, has been introduced to identify genuine quantum teleportation for which the general classical-teleportation model is intractable. It has been shown that this criterion is stricter than the existing criterion of $\bar{F}_{\rm{expt},\rm{s}}>0.667$ for surpassing the measure-prepare mimicry. We have additionally demonstrated that EPR steering powers genuine quantum teleportation to outperform classical teleportation, rather than entanglement. The proposed formalism determines, for the first time, the capability to mimic generic quantum dynamics using unsteerable states.

Regarding potential extensions of the proposed concept and method, future studies might usefully examine the applicability of the framework to quantum networks, which require the teleportation of genuinely multipartite entangled qubits for performing communication or computation \cite{Eisert00,Jiang07,Gottesman99,Chou18,Pirker18,Hsieh17,Chen20,Huang20}, where each qubit is transmitted via an individual teleportation channel. Such an investigation would be of benefit in determining how classical teleportation affects the characteristics of the teleported multi-qubit states. In particular, it is of significant interest for identifying the state resources required to perform networking quantum-information tasks which surpass classical mimicries. Furthermore, the proposed formalism may potentially be extended to the identification of quantum teleportation of more complex quantum systems, e.g., teleportation of multiple degrees of freedom \cite{Wang15} or teleportation in high dimensions \cite{Luo19,Hu19}.

\section*{Acknowledgements}
We thank Y.-A. Chen, Q. Zhang, H. Lu, and C.-Y. Lu for helpful comments and discussions. This work is partially supported by the Ministry of Science and Technology, Taiwan, under Grant Number MOST 107-2628-M-006-001-MY4.

\appendix

\section{Process matrix}

The Hermitian process matrix $\chi_{\rm{expt}}$ represents the linear teleportation completely positive map upon vectorisation of the density matrix \cite{Nielsen00,Breuer07}. The input states $\rho_{\rm{in}}$ and output states $\rho_{\rm{out}}$ of teleportation can be associated via the following dynamical mapping:
\begin{equation}
\chi_{\rm{expt}}:\rho_{\rm{in}}\mapsto \rho_{\rm{out}}.\nonumber
\end{equation}
That is, the output state can be explicitly represented as
\begin{equation}
\rho_{\rm{out}}\equiv\chi_{\rm{expt}}[\rho_{\rm{in}}]=\sum_{k=1}^{4}\sum_{j=1}^{4}\chi_{\rm{expt},kj}M_{k}\rho_{\rm{in}}M^{\dag}_{j},\label{rhoout}
\end{equation}
where $M_{k}=\ket{k_{1}}\!\!\bra{k_{2}}$, $M_{j}=\ket{j_{1}}\!\!\bra{j_{2}}$, $k=1+k_1+2k_2$, and $j=1+j_1+2j_2$ for $k_1,k_2,j_1,j_2\in\{0,1\}$. To determine the coefficients $\chi_{\rm{expt},kj}$ which constitute the process matrix $\chi_{\rm{expt}}$, let us consider the following four input states: $\rho_{\rm{in}}=\ket{0}\!\!\bra{0},\ket{0}\!\!\bra{1},\ket{1}\!\!\bra{0},\ket{1}\!\!\bra{1}$. From Eq.~(\ref{rhoout}), we obtain the corresponding output states as follows:
\begin{eqnarray}
\chi_{\rm{expt}}[\ket{0}\!\!\bra{0}]&=&\chi_{\rm{expt},11}\!\ket{0}\!\!\bra{0}+\chi_{\rm{expt},12}\ket{0}\!\!\bra{1}\nonumber\\
&& + \chi_{\rm{expt},21}\ket{1}\!\!\bra{0}+\chi_{\rm{expt},22}\ket{1}\!\!\bra{1}\nonumber\\
&=&\left[\begin{array}{cc}\chi_{\rm{expt},11} & \chi_{\rm{expt},12} \\ \chi_{\rm{expt},21} & \chi_{\rm{expt},22}\end{array}\right],\nonumber\\
\chi_{\rm{expt}}[\ket{0}\!\!\bra{1}]&=&\chi_{\rm{expt},13}\ket{0}\!\!\bra{0}+\chi_{\rm{expt},14}\ket{0}\!\!\bra{1}\nonumber\\
&& + \chi_{\rm{expt},23}\ket{1}\!\!\bra{0}+\chi_{\rm{expt},24}\ket{1}\!\!\bra{1}\nonumber\\
&=&\left[\begin{array}{cc}\chi_{\rm{expt},13} & \chi_{\rm{expt},14} \\ \chi_{\rm{expt},23} & \chi_{\rm{expt},24}\end{array}\right],\nonumber\\
\chi_{\rm{expt}}[\ket{1}\!\!\bra{0}]&=&\chi_{\rm{expt},31}\ket{0}\!\!\bra{0}+\chi_{\rm{expt},32}\ket{0}\!\!\bra{1}\nonumber\\
&& +\chi_{\rm{expt},41}\ket{1}\!\!\bra{0}+\chi_{\rm{expt},42}\ket{1}\!\!\bra{1}\nonumber\\
&=&\left[\begin{array}{cc}\chi_{\rm{expt},31} & \chi_{\rm{expt},32} \\ \chi_{\rm{expt},41} & \chi_{\rm{expt},42}\end{array}\right],\nonumber\\
\chi_{\rm{expt}}[\ket{1}\!\!\bra{1}]&=&\chi_{\rm{expt},33}\ket{0}\!\!\bra{0}+\chi_{\rm{expt},34}\ket{0}\!\!\bra{1}\nonumber\\
&& +\chi_{\rm{expt},43}\ket{1}\!\!\bra{0}+\chi_{\rm{expt},44}\ket{1}\!\!\bra{1}\nonumber\\
&=&\left[\begin{array}{cc}\chi_{\rm{expt},33} & \chi_{\rm{expt},34} \\ \chi_{\rm{expt},43} & \chi_{\rm{expt},44}\end{array}\right].\nonumber
\end{eqnarray}
It is clear that, once the density matrices of these four output states are known, the coefficients $\chi_{\rm{expt},kj}$ can be determined. One can use the method of quantum state tomography \cite{Nielsen00} to acquire a full knowledge of the output state density matrices $\chi_{\rm{expt}}[\ket{m}\!\!\bra{m}]=\rho_{\rm{out},m}$ for $m=0,1$. Moreover, since the coherence terms can be decomposed as $\ket{0}\!\!\bra{1}=\ket{+}\!\!\bra{+}\!+\!i\ket{R}\!\!\bra{R}\!-\!\hat{I} $ and $\ket{1}\!\!\bra{0}=\ket{+}\!\!\bra{+}\!-\!i\ket{R}\!\!\bra{R}\!-\!\hat{I}^{\dag}$, their output states are experimentally obtainable by measuring the density matrices of $\rho_{\rm{out},+}$ and $\rho_{\rm{out},R}$, i.e., $\chi_{\rm{expt}}[\ket{0}\!\!\bra{1}]= \rho_{\rm{out},+}+i\rho_{\rm{out},R}-\tilde{I}_{\rm{out}}$ and $\chi_{\rm{expt}}[\ket{1}\!\!\bra{0}]=\rho_{\rm{out},+}-i\rho_{\rm{out},R}-\tilde{I}_{\rm{out}}^{\dag}$. Therefore, with the above results, we arrive at Eq.~(\ref{exp}) and conclude that
\begin{eqnarray}
\chi_{\rm{expt}}
&=&\frac{1}{2}\left[\begin{array}{cccc}\chi_{\rm{expt},11} & \chi_{\rm{expt},12} & \chi_{\rm{expt},13} & \chi_{\rm{expt},14} \\\chi_{\rm{expt},21} & \chi_{\rm{expt},22} & \chi_{\rm{expt},23} & \chi_{\rm{expt},24} \\\chi_{\rm{expt},31} & \chi_{\rm{expt},32} & \chi_{\rm{expt},33} & \chi_{\rm{expt},34} \\\chi_{\rm{expt},41} & \chi_{\rm{expt},42} & \chi_{\rm{expt},43} & \chi_{\rm{expt},44}\end{array}\right]\nonumber\\
&=&\frac{1}{2}\left[\begin{array}{cc}\rho_{\rm{out},0} & \rho_{\rm{out},+}\!+\!i\rho_{\rm{out},R}\!-\!\tilde{I}_{\rm{out}} \\ \rho_{\rm{out},+}\!-\!i\rho_{\rm{out},R}\!-\!\tilde{I}_{\rm{out}}^{\dag} & \rho_{\rm{out},1}\end{array}\right],\nonumber\\
\end{eqnarray}
where the factor of $1/2$ is a normalization constant set such that $\chi_{\rm{expt}}$ can be treated as a density matrix.

\section{The average output state from the measure-prepare procedure}

In the measure-prepare strategy, Alice measures the unknown input state directly, and then sends the results via a classical communication channel to Bob to prepare the output state of teleportation \cite{Massar95}. Adopting the von Neumann measurement approach \cite{von55}, every measurement can be divided into two parts, namely the interaction between the two systems, i.e., the input state system and the measuring device system; and the manner in which the state of the measuring device is read. In order to acquire the information about the input state $\ket{m}$, Alice lets the input state interact with the measuring device, where the subsequent evolution of the input state takes place in accordance with an orthonormal basis denoted as $\{\ket{i}\}$. After each interaction round, Alice reads the final state of the measuring device and sends Bob a message via a classical channel to prepare the output state $\ket{i'}\in\{\ket{i}\}$. Given a sufficient number of rounds, the output state $\rho_m$ can be described by considering all the $\ket{i'}$ prepared by Bob in each round.

The interaction between the orthonormal basis $\ket{i}$ of the input state and the initial state of the measuring device $\ket{\phi_{0}}_{\text{\!MD}}$ can be described by
\begin{equation}\label{eq:mp00}
\ket{i}\ket{\phi_{0}}_\text{\!MD}
\rightarrow\sum_{f}
{\ket{f}{\ket{\phi^{i}_{f}}_\text{\!MD}}},
\end{equation}
where $\ket{f}$ are the final states of the input state $\ket{i}$ and are not necessarily orthogonal to each other or normalized. In addition, ${\ket{\phi^{i}_{f}}_{\text{\!MD}}}$ are the final states of the measuring device, which correspond to the evolution from $\ket{i}$ to $\ket{f}$ of the input-state system. The dimension of ${\ket{\phi^{i}_{f}}_{\text{\!MD}}}$ can be larger than or equal to that of the input state.

For general measurements, the final states of the measuring device ${\ket{\phi^{i}_{f}}_{\text{\!MD}}}$ need not be orthogonal to each other, nor normalized.
The only constraint is
\begin{equation}\label{eq:projector}
\sum_{f}
\tensor*[_{\text{MD\!}}]{\braket{
\phi^{i}_{f}|
{\phi}^{j}_{f}}}
{_{\text{\!MD}}}
=\delta_{ij},
\end{equation}
which ensures that the orthonormal basis $\ket{i}$ that interacted with the measuring device can be identified. For an arbitrary input state $\ket{m}$, the evolution of the total system is given as
\begin{equation}\label{eq:mp01}
\ket{m}{\ket{\phi_{0}}_{\text{\!MD}}}\rightarrow\sum_{i,f}
\braket{i|m}
\ket{f}{\ket{\phi^{i}_{f}}_{\text{\!MD}}}.
\end{equation}

After each round of interaction between the input state and measuring device,
Alice acquires the outcome by applying the one-dimensional projection operator
 $P_{{i'}\!f'}=
{\ket{\phi^{i'}_{f'}}_{\text{\!MD}}\!\!
\tensor*[_{\text{MD\!}}]{\bra{\phi^{i'}_{f'}}}{}}$
 onto the measuring device. To distinguish input states in the orthonormal basis $\ket{i}$ with the clearest distinction, the projectors $P_{{i'}\!f'}$ are orthogonal to each other for different $i'$ and $f'$. For the input state $\ket{m}$, the probability of measuring $\ket{\phi^{i'}_{f'}}_{\text{\!MD}}$ is
 \begin{equation}
 p_{i'\!,f'\!,m}\!\!=\!\!\!\sum_{i,i''\!,f,f''}\!\!\!
\braket{m|i}\!\!
\braket{i''|m}tr(\ket{f}\!\!\bra{f''}\otimes\ket{\phi^{i}_{f}}_{\text{\!MD}}\!\!\tensor*[_{\text{MD\!}}]{\bra{\phi^{i''}_{f''}}}{}\!
I\otimes P_{{i'}\!f'}).
\end{equation}
Since the projectors $P_{{i'}\!f'}$ are orthogonal to each other, i.e., $tr(P_{{i}f}P_{{i'}\!f'})$ $=tr(\ket{\phi^{i}_{f}}_{\text{\!MD}}\!\!
\tensor*[_{\text{MD\!}}]{\bra{\phi^{i}_{f}}}{}\ket{\phi^{i'}_{f'}}_{\text{\!MD}}\!\!
\tensor*[_{\text{MD\!}}]{\bra{\phi^{i'}_{f'}}}{})$
$=\delta_{ii'}\delta_{ff'}$, the probability $p_{i'\!,f'\!,m}$ can be re-written as
\begin{align}
 p_{i'\!,f'\!,m} &=&\sum_{i,i''\!\!,f,f''} &\braket{m|i}\!\!\braket{i''|m}\!\delta_{ii'}\!\delta_{i'i''}\delta_{ff'}\delta_{f'\!f''}\nonumber\\
&&& \ tr(\ket{f}\!\!\bra{f''}\!\otimes\!\ket{\phi^{i}_{f}}_{\text{\!MD}}\!\!\tensor*[_{\text{MD\!}}]{\bra{\phi^{i''}_{f''}}}{}
I\!\otimes\!P_{{i'}\!f'})\ \ \ \ \ \nonumber\\
&=&\braket{m|i'}&\!\! \braket{i'|m}\!tr(\ket{f'}\!\!\bra{f'}\!\otimes\!\ket{\phi^{i'}_{f'}}_{\text{\!MD}}\!\!\tensor*[_{\text{MD\!}}]{\bra{\phi^{i'}_{f'}}}{}
I\!\otimes\!P_{{i'}\!f'}).
\end{align}
According to Eq.~(\ref{eq:mp00}), $P_{{i'}\!f'}$ projects the state of the measuring device onto the final state ${\ket{\phi^{i'}_{f'}}_{\text{\!MD}}}$,
 which describes the evolution from $\ket{i'}$ to $\ket{f'}$ of the input-state system. Alice then asks Bob to prepare state $\ket{i'}$.
 After a sufficient number of runs,
 the output state $\rho_{m}$ can be described by
\begin{equation}\label{eq:mp03}
\rho_m
=
\sum_{i',f'}
p_{i'\!,f'\!,m}
\ket{i'}\!\bra{i'}.
\end{equation}
Since the projectors $P_{{i'}\!f'}$ form a complete set of orthogonal projectors, if the input state is $\ket{i}$, the sum of probabilities for different results is one, i.e., $\sum_{i'\!,f,f'\!,f''}\delta_{ii'}\delta_{ff'}\delta_{f'\!f''}tr(\ket{f}\!\!\bra{f''}\!\otimes\!\ket{\phi^{i}_{f}}_{\text{\!MD}}\!\!\tensor*[_{\text{MD\!}}]{\bra{\phi^{i}_{f''}}}{}
I\!\otimes\!P_{{i'}\!f'})=\sum_{f'}tr(\ket{f'}\!\!\bra{f'}\!\otimes\!\ket{\phi^{i'}_{f'}}_{\text{\!MD}}\!\!\tensor*[_{\text{MD\!}}]{\bra{\phi^{i'}_{f'}}}{}
I\!\otimes\!P_{{i'}\!f'})=1$, and the output state $\rho_m$ can be re-written as
\begin{eqnarray}\label{eq:mp02}
&&\rho_m\nonumber\\
&&=\!\!\sum_{i'\!,f'}\!\!\braket{m|i'}\!\!
\braket{i'|m}\!tr(\ket{f'}\!\!\bra{f'}\!\otimes\!\ket{\phi^{i'}_{f'\!}}_{\text{\!MD}}\!\!\!\tensor*[_{\text{MD\!}}]{\bra{\phi^{i'}_{f'\!}}}{}\!
I\!\otimes\!P_{{i'}\!f'}\!)\!\ket{i'}\!\!\bra{i'}\ \ \ \ \ \nonumber \\
&&=\!\!\sum_{i'}\!\!\braket{m|i'}\!\!
\braket{i'|m}\!\!\sum_{f'}\!tr(\ket{f'}\!\!\bra{f'}\!\otimes\!\ket{\phi^{i'}_{f'\!}}_{\text{\!MD}}\!\!\!\tensor*[_{\text{MD\!}}]{\bra{\phi^{i'}_{f'\!}}}{}\!
I\!\otimes\!P_{{i'}\!f'}\!)\!\ket{i'}\!\!\bra{i'}\nonumber \\
&&=\!\!\sum_{i'}p_{i'\!,m}\ket{i'}\!\!\bra{i'},
\end{eqnarray}
where $p_{i'\!,m}=\braket{m|i'}\!\!
\braket{i'|m}=|\!\left\langle i'|m\right\rangle\!|^{2}$.

The state~(\ref{eq:mp02}) corresponds to Eq.~(\ref{rm}) for measuring only one observable, say $A_{k'}$. See also~Fig.~\ref{QT}(b). In general, Alice can use more than one measurement device to measure the input state $\ket{m}$ with respect to different observables $A_{k}$ with eigenbases $\{\ket{a_k}\}$. Thus, the output state $\rho_m$ can be represented by Eq.~(\ref{rm}) where $p_{k}$ is the probability that Alice chooses the measurement device for measuring $A_{k}$.

\section{Extended measure-prepare model described by~(\ref{unsteerable}) and~(\ref{outputs})}

In this section, we show that Eqs.~(\ref{p_xi_solution}), (\ref{recipe1}) and (\ref{P_lambda_A_xi_a})--(\ref{rho_B_xi_m}) can describe the extended measure-prepare model proposed in Ref. \cite{Cavalcanti17} using our classical model in~(\ref{unsteerable}) and~(\ref{outputs}).

\noindent (i) Product measurement. We show that the outputs states of the case using product measurement in the extended measure-prepare model [Eqs.~(\ref{rho_out_m_product}) and~(\ref{rho_out_m_a_product})] can be described by Eqs.~(\ref{p_xi_solution}) and~(\ref{recipe1}) according to Eqs.~(\ref{unsteerable}) and~(\ref{outputs}). From the output states of classical teleportation~[Eqs.~(\ref{unsteerable}) and~(\ref{outputs})] and the extended measure-prepare model~[Eqs.~(\ref{rho_out_m_product}) and~(\ref{rho_out_m_a_product})] $\rho_{\text{out},m}$ for $m=0,$ $1,$ $+,$ $-,$ $R$ and~$L$, the following system of six bilinear equations can be obtained:
\begin{eqnarray}
\rho_{\text{out},0}&=&\sum_{\xi=1,3,5,7}\!\!2P(\lambda^{A}_{\xi}){\rho^{B}_{\xi}}=\sum_{a}P(a|0)U_a\rho^{B}_{a}U_a^\dagger,\nonumber\\
\rho_{\text{out},1}&=&\sum_{\xi=2,4,6,8}\!\!2P(\lambda^{A}_{\xi}){\rho^{B}_{\xi}}=\sum_{a}P(a|1)U_a\rho^{B}_{a}U_a^\dagger,\nonumber\\
\rho_{\text{out},+}&=&\sum_{\xi=1,2,3,4}\!\!2P(\lambda^{A}_{\xi}){\rho^{B}_{\xi}}=\sum_{a}P(a|+)U_a\rho^{B}_{a}U_a^\dagger,\nonumber\\
\rho_{\text{out},-}&=&\sum_{\xi=5,6,7,8}\!\!2P(\lambda^{A}_{\xi}){\rho^{B}_{\xi}}=\sum_{a}P(a|-)U_a\rho^{B}_{a}U_a^\dagger,\nonumber\\
\rho_{\text{out},R}&=&\sum_{\xi=1,2,5,6}\!\!2P(\lambda^{A}_{\xi}){\rho^{B}_{\xi}}=\sum_{a}P(a|R)U_a\rho^{B}_{a}U_a^\dagger,\nonumber\\
\rho_{\text{out},L}&=&\!\!\sum_{\xi=3,4,7,8}\!\!2P(\lambda^{A}_{\xi}){\rho^{B}_{\xi}}=\sum_{a}P(a|L)U_a\rho^{B}_{a}U_a^\dagger.\label{eq_solved}
\end{eqnarray}
There are 16 variables, $P(\lambda^{A}_{\xi})$ and $\rho^{B}_{\xi}$ for $\xi=1,2,...,8$, in Eq~(\ref{eq_solved}).
By assuming $P(\lambda^{A}_{\xi})=1/8$ for $\xi=1,2,...,8$ (\ref{p_xi_solution}), Eq.~(\ref{eq_solved}) becomes a system of linear equations involving 8 variables and thus has more than one solution.
To solve $\rho^{B}_{\xi}$, one can sum specific equations in (\ref{eq_solved}). For example, for $\rho^{B}_{1}$, which represents the corresponding particle state of the pre-existing state $\lambda^{A}_{1}(+1,+1,+1)$ (\ref{unsteerable}) and is a constituent of $\rho_{\text{out},0}$, $\rho_{\text{out},+}$ and $\rho_{\text{out},R}$ (\ref{eq_solved}), one can sum the equations of $\rho_{\text{out},0}$, $\rho_{\text{out},+}$, and $\rho_{\text{out},R}$ and obtain
\begin{eqnarray}
&&\rho_{\text{out},0}+\rho_{\text{out},+}+\rho_{\text{out},R}\nonumber\\
&&=(\sum_{\xi=1,3,5,7}\frac{1}{4}{\rho^{B}_{\xi}})+(\sum_{\xi=1,2,3,4}\frac{1}{4}{\rho^{B}_{\xi}})+(\sum_{\xi=1,2,5,6}\frac{1}{4}{\rho^{B}_{\xi}})\nonumber\\
&&=\frac{1}{4}(3{\rho^{B}_{1}}\!+\!2{\rho^{B}_{2}}\!+\!2{\rho^{B}_{3}}\!+\!{\rho^{B}_{4}}\!+\!2{\rho^{B}_{5}}\!+\!{\rho^{B}_{6}}\!+\!{\rho^{B}_{7}}).\label{out_0pr}
\end{eqnarray}
From Eq.~(\ref{unsteerable}), $(\rho^{B}_{2}+\rho^{B}_{7})/2$ describes the output state of the input state $I/2$, denoted as $I_{\text{out}}/2$, since the measurement outcomes of $I/2$ can be described by the equal mixture of the two pre-existing states, $\lambda^{A}_{2}$ and $\lambda^{A}_{7}$. Similarly, $(\rho^{B}_{3}+\rho^{B}_{6})/2=(\rho^{B}_{4}+\rho^{B}_{5})/2=I_{\text{out}}/2$. From the pre-existing states in (\ref{unsteerable}), $(\rho^{B}_{2}+\rho^{B}_{3}+\rho^{B}_{5})/3$ describes the output of one-third $\lambda^{A}_{1}$ and two-thirds $I/2$, i.e.,
\begin{eqnarray}
{\frac{\rho^{B}_{2}+\rho^{B}_{3}+\rho^{B}_{5}}{3}=\frac{1}{3}\rho^{B}_{1}+\frac{2}{3}\frac{I_{\text{out}}}{2}}.\nonumber
\end{eqnarray}
$\rho_{\text{out},0}+\rho_{\text{out},+}+\rho_{\text{out},R}$ {in (\ref{out_0pr})} can then be re-written as
\begin{eqnarray}
&&\rho_{\text{out},0}\!+\!\rho_{\text{out},+}\!+\!\rho_{\text{out},R}\nonumber\\
&&=\frac{1}{4}[3{\rho^{B}_{1}}\!+\!({\rho^{B}_{2}}\!+\!{\rho^{B}_{7}})\!+\!({\rho^{B}_{3}}\!+\!{\rho^{B}_{6}})\!+\!({\rho^{B}_{4}}\!+\!{\rho^{B}_{5}})\!+\!({\rho^{B}_{2}}\!+\!{\rho^{B}_{3}}\!+\!{\rho^{B}_{5}})]\nonumber\\
&&={\rho^{B}_{1}}\!+\!{2\frac{I_{\text{out}}}{2}}.\nonumber
\end{eqnarray}
{From Eq.~(\ref{rho_out_m_product}),} {since $I/2=(\ket{0}\!\!\bra{0}+\ket{1}\!\!\bra{1})/2$, ${I_{\text{out}}}/{2}$ is}
\begin{eqnarray}
{\frac{I_{\text{out}}}{2}}&=&{\frac{1}{2}(\rho_{\text{out},0}+\rho_{\text{out},1})} \nonumber \\
&=&\frac{1}{2}\sum_{a}[P(a|0)+P(a|1)]U_a\rho^{B}_{a}U_a^\dagger \nonumber \\
&=&\frac{1}{2}\sum_{a}\left[\frac{P(a,0)}{P(0)}+\frac{P(a,1)}{P(1)}\right]U_a\rho^{B}_{a}U_a^\dagger \nonumber \\
&=&\sum_{a}P(a)U_a\rho^{B}_{a}U_a^\dagger,\nonumber
\end{eqnarray}
{under the assumption of a normal distribution of input $m$, i.e., $P(m)=1/2\ \ \forall m$.}
Thus, the solution of $\rho^{B}_{1}$ (\ref{recipe1}) is obtained through
\begin{eqnarray}
&&\rho_{\text{out},0}+\rho_{\text{out},+}+\rho_{\text{out},R}\nonumber\\
&&={\rho^{B}_{1}}+2\sum_{a}P(a)U_a\rho^{B}_{a}U_a^\dagger\nonumber\\
&&=\sum_{a}[P(a|0)+P(a|+)+P(a|R)]U_a\rho^{B}_{a}U_a^\dagger\nonumber
\end{eqnarray}
and one gets
\begin{eqnarray}
\rho^{B}_{1}=\sum_{a}[P(a|0)+P(a|+)+P(a|R)-2P(a)]U_a\rho^{B}_{a}U_a^\dagger.\nonumber
\end{eqnarray}
The solution of $\rho_{\xi}^{B}$ in (\ref{recipe1}) can be obtained in a similar fashion.\\

\noindent (ii) Bell-state measurement. For the case where the Alice's measurements are partial Bell-state measurements, we first present the derivation of Eqs.~(\ref{P_lambda_A_xi_a}) and~(\ref{rho_B_xi_a}), i.e., the $P(\lambda^{A}_{\xi}|a')$ and $\rho^{B}_{\xi,a'}$ used to describe the state $\rho_{\text{out},m,a'}$ (\ref{noroutput}). We then show that Eq.~(\ref{rho_out_m_U_a}) can be represented in the form of (\ref{unsteerable}) and~(\ref{outputs}) with $P({\lambda'}^{A}_{\xi})$ (\ref{P_lambda_A_xi_m}) and ${\rho'}^{B}_{\xi}$ (\ref{rho_B_xi_m}).

To describe the output states $\rho_{\text{out},m,a'}$ (\ref{noroutput}) with the pre-existing states $\lambda^{A}_{\xi}$ and ${\rho}^{B}_{\xi}$, we first rewrite the conditional probability $P(m|\kappa,a')$ in (\ref{noroutput}) as
\begin{eqnarray}
P(m|\kappa,a')&=&\sum_{\xi}P(m,\lambda^{A}_{\xi}|\kappa,a')\nonumber \\
&=&\sum_{\xi}P(\lambda^{A}_{\xi}|\kappa,a')P(m|\lambda^{A}_{\xi},\kappa,a').\nonumber
\end{eqnarray}
According to Eq.~(\ref{unsteerable}), conditioned on a specific $m=m'$, $P(m|\lambda^{A}_{\xi},\kappa,a')$ is nonzero for specific $\xi$. Thus, $P(m'|\kappa,a')$ becomes
\begin{eqnarray}
P(m'|\kappa,a')=\sum_{\xi|m'}P(\lambda^{A}_{\xi}|\kappa,a').\label{m_kappa}
\end{eqnarray}
Substituting Eq.~(\ref{m_kappa}) into Eq.~(\ref{noroutput}), $\rho_{\text{out},m,a'}$ can be re-written as
\begin{eqnarray}
\rho_{\text{out},m,a'}&=&\frac{1}{P(m|a')}\sum_{\xi|m}\sum_{\kappa}p_{\kappa}P(\lambda^{A}_{\xi}|\kappa,a')\rho^{B}_{\kappa}\nonumber \\
&=&\sum_{\xi|m}\sum_{\kappa}2p_{\kappa}P(\lambda^{A}_{\xi}|\kappa,a')\rho^{B}_{\kappa},\label{output_1}
\end{eqnarray}
under the assumption of a normal distribution of the input states, i.e., $P(m|a')=1/2$.
We then define $\rho^{B}_{\xi,a'}$ which is in the form shown in (\ref{rho_B_xi_a}), and $P(\lambda^{A}_{\xi}|a')=\sum_{\kappa}P(\lambda^{A}_{\xi},\kappa|a')$.
The output states $\rho_{\text{out},m,a'}$ in Eq.~(\ref{output_1}) can then be re-written as
\begin{eqnarray}
\rho_{\text{out},m,a'}=\sum_{\xi|m}2P(\lambda^{A}_{\xi}|a')\rho^{B}_{\xi,a'}.\nonumber
\end{eqnarray}
Regarding the input states for QPT, i.e., $m=0,$ $1,$ $+,$ and $R$, the corresponding output states $\rho_{\text{out},0,a'},$ $\rho_{\text{out},1,a'},$ $\rho_{\text{out},+,a'},$ and $\rho_{\text{out},R,a'}$ are given as
\begin{eqnarray}
&& \rho_{\text{out},0,a'}=\sum_{\xi=1,3,5,7}2P({\lambda^{A}_\xi}|a'){\rho^{B}_{\xi,a'}},\nonumber\\
&&  \rho_{\text{out},1,a'}=\sum_{\xi=2,4,6,8}2P({\lambda^{A}_\xi}|a'){\rho^{B}_{\xi,a'}},\nonumber\\
&& \rho_{\text{out},+,a'}=\sum_{\xi=1,2,3,4}2P({\lambda^{A}_\xi}|a'){\rho^{B}_{\xi,a'}},\nonumber\\
&&  \rho_{\text{out},R,a'}=\sum_{\xi=1,2,5,6}2P({\lambda^{A}_\xi}|a'){\rho^{B}_{\xi,a'}},\nonumber
\end{eqnarray}
and can be described by our classical-teleportation model in~(\ref{unsteerable}) and (\ref{outputs}).

Considering the correction $U_a$ for each case $a$ in Eq.~({\ref{out_m}}), the output states $\rho_{\text{out},m}$ become the form shown in (\ref{rho_out_m_U_a})
and consist of the pre-existing recipe for classical teleportation used in each case $a$, i.e., $P(\lambda^{A}_{\xi}|a)$ and $\rho^{B}_{\xi,a}$. Since $P(\lambda^{A}_{\xi},a)=P(\lambda^{A}_{\xi})P(a|\lambda^{A}_{\xi})$, through the definitions of $P({\lambda}^{A}_{\xi})$ and ${\rho}^{B}_{\xi}$ in (\ref{P_lambda_A_xi_m}) and (\ref{rho_B_xi_m}), respectively, the output states $\rho_{\text{out},m}$ in (\ref{rho_out_m_U_a}) become
\begin{eqnarray}
\rho_{\text{out},m}&=&\sum_{\xi|m}2P(\lambda^{A}_{\xi})\sum_{a}P(a|\lambda^{A}_{\xi})U_a\rho^{B}_{\xi,a}U_a^\dagger \nonumber\\
&=&\sum_{\xi|m}2P(\lambda^{A}_{\xi})\rho^{B}_{\xi},\label{output_mp}
\end{eqnarray}
which has the same form as Eq.~(\ref{outputs}) and shows that the extended measure-prepare model is a special case of classical teleportation.

\section{Proof of $F_{\text{{s}}}=F_{\text{expt}}$ in Eq.~(\ref{sfidelity})}

We assume that the EPR pair of the state $\ket{\phi^+}$ for teleportation [Fig.~\ref{QT}(a)] is replaced by a pair with the state $\rho_{\text{expt}}$. To consider how the resulting teleportation depends on the proportion of $\ket{\phi^+}$ in $\rho_{\text{expt}}$, we first represent $\rho_{\text{expt}}$ in the basis consisting of the outer products of the Bell states, and arrive at
\begin{equation}
\rho_{\text{expt}}=\sum_{b,d=\phi^\pm,\psi^\pm}f_{bd}\ket{b}\!\!\bra{d},\label{rexpt}
\end{equation}
where $f_{bd}$ are the entries of the density matrix $\rho_{\text{expt}}$. Note that the state fidelity $F_{\text{{s}}}$ in Eq.~(\ref{sfidelity}) is equal to the coefficient $f_{\phi^+\!\phi^+}$.

Ideally, as $\ket{\phi^+}$ is used to teleport the input state $\rho_{\text{in},m}$, the whole process acts as an identity operation (see Eq.~(\ref{i})). By contrast, when $\ket{\phi^+}$ is replaced with the other Bell states: $\ket{\phi^-}$, $\ket{\psi^+}$ and $\ket{\psi^-}$, and teleportation is assumed to be implemented using the same protocol designed for $\ket{\phi^+}$, the corresponding output states become: $Z\rho_{\text{in},m}Z^{\dag}$, $X\rho_{\text{in},m}X^{\dag}$ and $Y\rho_{\text{in},m}Y^{\dag}$, respectively. Therefore, when $\rho_{\text{expt}}$ (\ref{rexpt}) is used for teleportation, we obtain the following output state:
\begin{equation}
\rho_{\text{out},m}=\chi_{\text{expt}}[\rho_{\text{in},m}]=\sum_{b,d=\phi^\pm,\psi^\pm}f_{bd}E_{b}\rho_{\text{in},m}E^{\dag}_{d},\label{rhomout}
\end{equation}
where $E_{\phi^+}=I$, $E_{\phi^-}=Z$, $E_{\psi^+}=X$ and $E_{\psi^-}=Y$. Eq.~(\ref{rhomout}) shows that the coefficients $f_{ab}$ constitute the process matrix of $\chi_{\text{expt}}$ in the basis of the Pauli matrices $\{I,X,Y,Z\}$, rather than the normal basis used in Eq.~(\ref{rhoout}). In other words, $f_{\phi^+\!\phi^+}$ can be considered as the process fidelity of experimental teleportation $\chi_{\text{expt}}$ and ideal teleportation, $\chi_{\text{I}}$, $F_{\text{expt}}$. We therefore conclude that $F_{\text{{s}}}=F_{\text{expt}}$.

\section{Example of classical teleportation $\chi_{\text{CT}}$ in Fig.~\ref{example}}

Consider the example in Fig.~\ref{example} in the main text, which shows $F_{\text{expt}}$ for the case where the entanglement source of a teleportation experiment mixes with white noise and becomes $\rho_\text{{W}}(p_{\text{noise}})$ (see Eq.~(\ref{rw})). In the range $0.423 \leq p_{\text{noise}} < 0.667$, where $0.5 < F_{\text{expt}}\leq 0.683$, the teleportation process, $\chi_{\text{expt}}(p_{\text{noise}})$, cannot be simulated by the measure-prepare strategy, but can still be described by the classical-teleportation model, $\chi_{\text{CT}}$. For example, given a white noise intensity of $p_{\text{noise}}=0.5$, the teleportation process with $F_{\text{expt}}=0.625$:
\begin{eqnarray}
\chi_{\text{expt}}(0.5)= \left[ \begin{array}{cccc}
    0.375 & 0 & 0 & 0.250\\
    0 & 0.125 & 0 & 0\\
    0 & 0 & 0.125 & 0\\
    0.250 & 0 & 0 & 0.375
    \end{array}
\right],\label{chi_expt}
\end{eqnarray}
can be described by $\chi_{\text{CT}}$ with $P(\lambda^{A}_{\xi})=1/8$ for $\xi=1,2,...,8$,

\begin{widetext}
\begin{eqnarray}
&&\rho^{B}_{1}= \left[ \begin{array}{cc}
    0.75 & 0.25(1-i) \\
    0.25(1+i) & 0.25
    \end{array}
\right],
\rho^{B}_{2}= \left[ \begin{array}{cc}
    0.25 & 0.25(1-i) \\
    0.25(1+i) & 0.75
    \end{array}
\right],
\rho^{B}_{3}= \left[ \begin{array}{cc}
    0.75 & 0.25(1+i) \\
    0.25(1-i) & 0.25
    \end{array}
\right],\nonumber\\
&&\rho^{B}_{4}= \left[ \begin{array}{cc}
    0.25 & 0.25(1+i) \\
    0.25(1-i) & 0.75
    \end{array}
\right],
\rho^{B}_{5}= \left[ \begin{array}{cc}
    0.75 & -0.25(1+i) \\
    -0.25(1-i) & 0.25
    \end{array}
\right],
\rho^{B}_{6}= \left[ \begin{array}{cc}
    0.25 & -0.25(1+i) \\
    -0.25(1-i) & 0.75
    \end{array}
\right],\nonumber\\
&&\rho^{B}_{7}= \left[ \begin{array}{cc}
    0.75 & -0.25(1-i) \\
    -0.25(1+i) & 0.25
    \end{array}
\right],
\rho^{B}_{8}= \left[ \begin{array}{cc}
    0.25 & -0.25(1-i) \\
    -0.25(1+i) & 0.75
    \end{array}
\right].\nonumber
\end{eqnarray}
\end{widetext}
See Eqs.~(\ref{unsteerable}) and~(\ref{ct}) in the main text for details.

\end{document}